\documentclass[10pt,a4paper]{article}
\usepackage{graphpap,epsfig,amssymb,graphicx,textcomp}
\usepackage[utf8]{inputenc}
\usepackage[english]{babel}
\begin{document}
\begin{center}{\Large{\bf Transient phases in the Vicsek model of flocking}}
\end{center}

\vskip 1cm

\begin{center}{\it {Sayantani Kayal$^1$ and Muktish Acharyya$^2$}}\\
{\it Department of Physics, Presidency University}\\
{\it 86/1 College street, Kolkata-700073, India}\\
{E-mail$^1$: iamsayantanikayal@gmail.com}\\
{E-mail$^2$: muktish.physics@presiuniv.ac.in}\end{center}

\vskip 2cm

\noindent {\bf Abstract:} The Vicsek model of flocking is studied by computer
simulation. We confined our studies here to the morphologies and the
lifetimes of transient phases. In our simulation, we have identified three distinct
transient phases, namely, vortex phase, colliding phase and multi-domain phase. The
mapping of Vicsek model to XY model in the $v \to 0$ limit prompted us to explore
the possibility of finding any vortex kind of phases in the Vicsek model. We have
obtained rotating vortex phase and measured the lifetime of this vortex phase. We
have also measured the lifetimes of other two transient 
phases, i.e., colliding phase and multi-domain phase. We have measured the integrated lifetime
($\tau_t$) of all these transient phases
and studied this as function of density ($\rho$) and noise ($\eta$).
In the low noise regime, we proposed here a scaling law $\tau_t N^{-a} = F(\rho
N^{-b})$ where $F(x)$ is a scaling function like $F(x) \sim x^{-s}$. By the method
of data collapse, we have estimated the exponents as $a=-0.110\pm0.010$, 
$b=0.950\pm0.010$
and $s=1.027\pm0.008$. The integrated lifetime $\tau$
(defined in the text differently) was observed to decrease
as the noise approaches the critical noise from below. 
This behaviour is quite unusual and contrary to the critical slowing down observed
in the case of well-known equilibrium phase transitions.
We have provided a possible
explanation from the time evolution of the distribution of the directions of
velocities.

\vskip 1cm


\vskip 1cm

\newpage

\noindent {\bf 1. Introduction:}

\vskip 0.1cm 

The non-equilibrium phase transition under driven noise, shown by the
natural systems possessing collective behaviour, is being studied
\cite{sriram1,sriram2,sriram3} with
much interest.
In nature, we find the collective motion of large groups of motile agents
(e.g., collective motion of animal groups\cite{jtb}, flocking of birds, schooling of fish\cite{fish}, motion
of bacteria and microorganisms\cite{cell} etc.) which move over a distance much larger 
than the size
of the motile agents. Despite the difference in microscopic interactions
of different systems, 
such collective motions 
can be described by the common macroscopic behaviours
of simple physical models. In 1995, T. Vicsek and coworkers
proposed\cite{vicsek}
 such a model, which consists of point particles moving with constant
absolute velocity but imperfectly aligned in the direction with their
neighbours. In the limit of low imperfection in alignment, a phase
transition is observed, where all agents move in a particular direction
coherently. Needless to say, for higher values of noise, such kind of
unidirectional coherent motion is not found.
This flocking transition gives rise to symmetry broken ordered
phase with long range order even in two dimensions (which was forbidden
at equilibrium by Mermin and Wagner theorem\cite{mermin}). This is the
key reason of interest taken by the modern researchers to study
\cite{revvicsek,ginelli} the
Vicsek model of flocking. A considerable amount of investigations were
made in recent years which prompted intense interests of active research
in Physics\cite{filella} even today. We briefly mention a few of those here. 

A variant of original Vicsek model was studied\cite{manna1} with binary
interactions among the flocks and a topological distance dependent transition
is observed. They\cite{manna2} have also observed coherent
and interesting cyclic
states (circular motion of the agents) with topological distances.
Very recently, the original Vicsek model is modified\cite{arnab} by 
introducing the time delay
in the interaction and angular restriction (vision cone). 
This modification leads to
a state where the agents spontaneously condense into a dense drop.
The effects of inertia (microorganisms having finite size and mass)
of the model flocks in a turbulent environment 
are also studied\cite{samriddhi} recently.

A considerable amount of work has been done
 to study the nature of the phases and the transitions
(continuous or discontinuous)
in Vicsek model of flocking. It is reported\cite{solon1} that the phase 
transition in the Vicsek model is best understood as a liquid-gas like 
transition (which shows metastability, hysteresis etc.) which is in contrast
to the bulk phase separation observed\cite{solon2} in active Ising model.
From the hydrodynamic equations,
three different kinds of solutions are obtained\cite{solon3}, namely,
periodic orbits (microphase separation), homoclinic orbits (solitonic
objects) and heteroclinic trajectories (phase separation).
Although, the scaling was proposed by Vicsek similar to equilibrium
critical phenomena,
later, it was shown\cite{chate1} that this transition is indeed first order.

Since the Vicsek model can provide a special non-equilibrium phase transition
which breaks the symmetry in ordered phase even in two dimensions, it would be
interesting to know whether any vortex kind of phase can be observed in the
limit of $v \to 0$\cite{vicsek}. This may be a common question, what kind of
transient phases can be observed, in general.
In this paper, we have systematically studied the temporal evolution
of the transient phases 
in the Vicsek model and their lifetimes as function of density and 
noise. The paper is organised as follows: In section-2 the Vicsek model is
described, in section-3 the methodology and the numerical results are
reported and the paper ends with a summary in section-4.\\

\vskip 0.5cm

\noindent {\bf 2. The Vicsek model:}
\vskip 0.1cm
The original Vicsek model of flocking is described as follows: identical point-wise particles move continuously on an off-lattice plane represented by a two dimensional square shaped cell of linear size $ L $ with periodic boundary conditions. At $ t $=0, $ N $ number of particles (i) were randomly distributed in the cell with the position of $i^{th}$ particle denoted by $ \lbrace x_{i},y_{i} \rbrace $, (ii) had the same absolute velocity $ v $ and (iii) had randomly distributed directions with the direction of $i^{th}$ particle denoted by $ \theta_{i} $. The particles move in discrete time steps, i.e., the time interval ($ \Delta t $) between two updating of directions and positions is considered to be unity ($\Delta t =1$). At each time step, the velocities $ \lbrace vcos\theta_{i},vsin\theta_{i} \rbrace $ of the particles are determined simultaneously (by parallel updating of all the particles) and the position of the $i^{th}$ particle is updated according to the relations
\begin{equation}
x_{i}(t+1)=x_{i}(t)+vcos\theta_{i}(t)
\end{equation}
\begin{equation}
y_{i}(t+1)=y_{i}(t)+vsin\theta_{i}(t)
\end{equation}
and the angle is obtained from the expression
\begin{equation}
\theta_{i}(t+1)=\langle\theta_{i}(t)\rangle_{r}+\Delta \theta
\end{equation}
where $\langle\theta_{i}(t)\rangle_{r}$ denotes the average direction of the velocities of the particles (including the $i^{th}$ particle) surrounding the given particle within a circle of radius $r$. The average direction is given by the angle $tan^{-1}[\langle sin\textbf{(}\theta_{i}(t)\textbf{)} \rangle_{r} / \langle cos\textbf{(}\theta_{i}(t)\textbf{)} \rangle_{r}]$ and $\Delta \theta$ is a random number chosen with a uniform probability from the interval [$-\eta /2,\eta /2$] responsible for the noise in the velocity direction. 
This noise plays the role of indeterminacy in choosing the direction of motion. The noise tends to 
prevent the ordering in the system.
$\Delta \theta$ is delta-correlated scalar \textit{white} noise ranging maximally 
in the interval [-$\pi,\pi$]. The equation (3) shows that each particle tends to self-organize in the same average direction of motion (first term) while the behaviour being randomly perturbed (second term). The flock synchronously evolves through an iteration of these rules and that is studied in detail by determining the absolute value of the average normalised velocity 
\begin{equation}
v_{a} = {1 \over Nv} \left| \sum_{i=1}^{N} \mathbf{v_{i}} \right|
\end{equation} 
of the entire system of particles. The velocity $v_{a}$ is approximately $0$ if the direction of motion of individual particles is distributed randomly (with uniform probability), while for the case of coherent motion (with ordered direction of velocities), $v_{a}\cong$ 1. Here, the average normalised velocity is considered as an \textit{order parameter} of the system. In the steady state, the system undergoes a kinetic phase transition from no transport to finite net transport as the noise ($\eta$) and density ($\rho = N/L^{2}$) of the system are tuned individually. The critical value of noise corresponding to the phase transition is observed in this study by the singular nature (sharply peaked for finite system) of the variance of order parameter in the critical region. This variance (or, standard deviation) has played important role throughout this study in determining several properties of various phases which is to be discussed in the subsequent sections.\\

\vskip 0.5cm

\noindent {\bf 3. Simulations and results:}
\vskip 0.1cm
Inspired by the anticipated analogy of Vicsek model in the $v\rightarrow 0$ limit to the well-known XY-model, we initially speculated possible existence of vortices in the low velocity limit corresponding to analogous Kosterlitz-Thouless transition\cite{koster}. 
However, even while working with $v=0.03$ which is used by Vicsek et. al. to study the kinetic phase transition, we have observed not only metastable vortices, but also two other novel transient phases with interesting dynamics. We have estimated their properties and lifetimes individually and have studied the dependence of the integrated lifetime of all these transient phases
 on density ($\rho$) and noise ($\eta$) of the system. We have explained some novel phenomena obtained in this study with the help of standard statistical tools and techniques.

Throughout all our simulations, the absolute velocity $v$ and the radius of influence $r$ are kept fixed at values $v=0.03$ and $r=1.0$ respectively. Except for the cases reported in section 3.1, all the characterisations of the transient phases are done using 1000 ensembles of the single realisation.\\

\vskip 0.3cm

\begin{figure}[h]
\begin{center}
\begin{tabular}{c}
\resizebox{7.0cm}{!}{\includegraphics[angle=0]{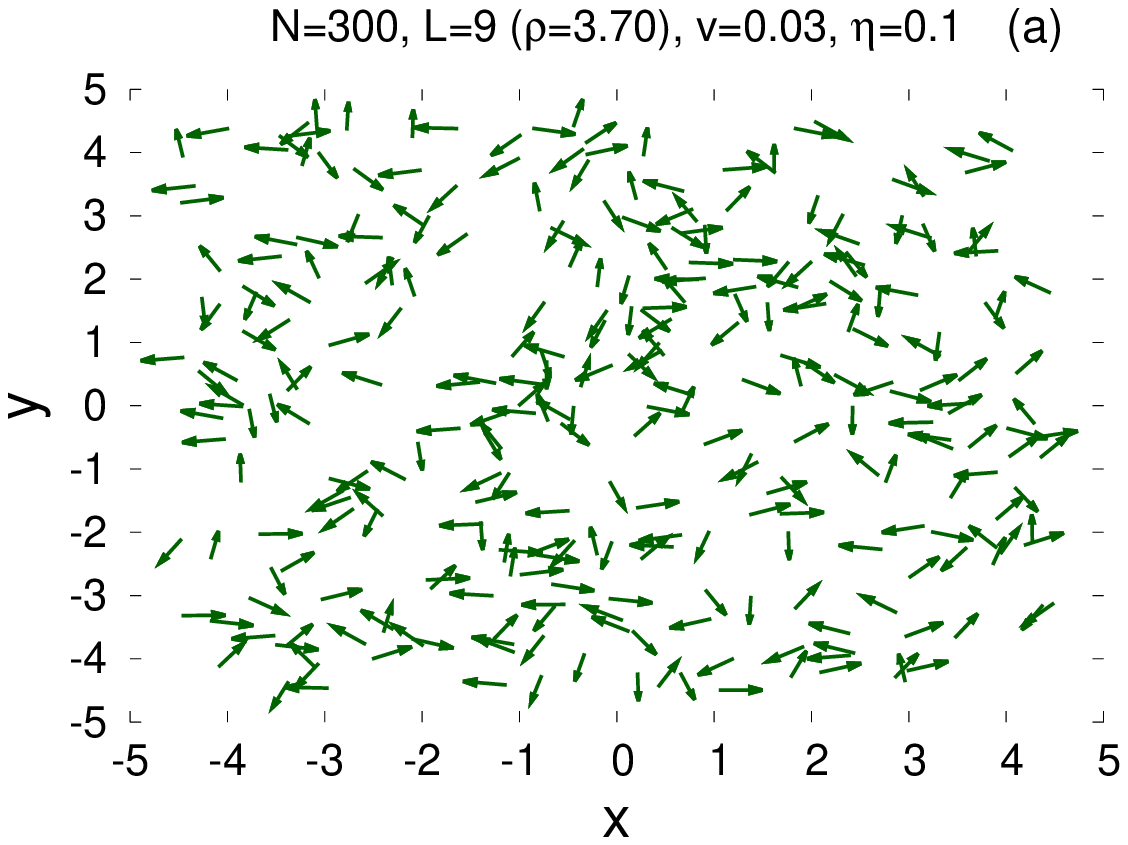}}
\resizebox{7.0cm}{!}{\includegraphics[angle=0]{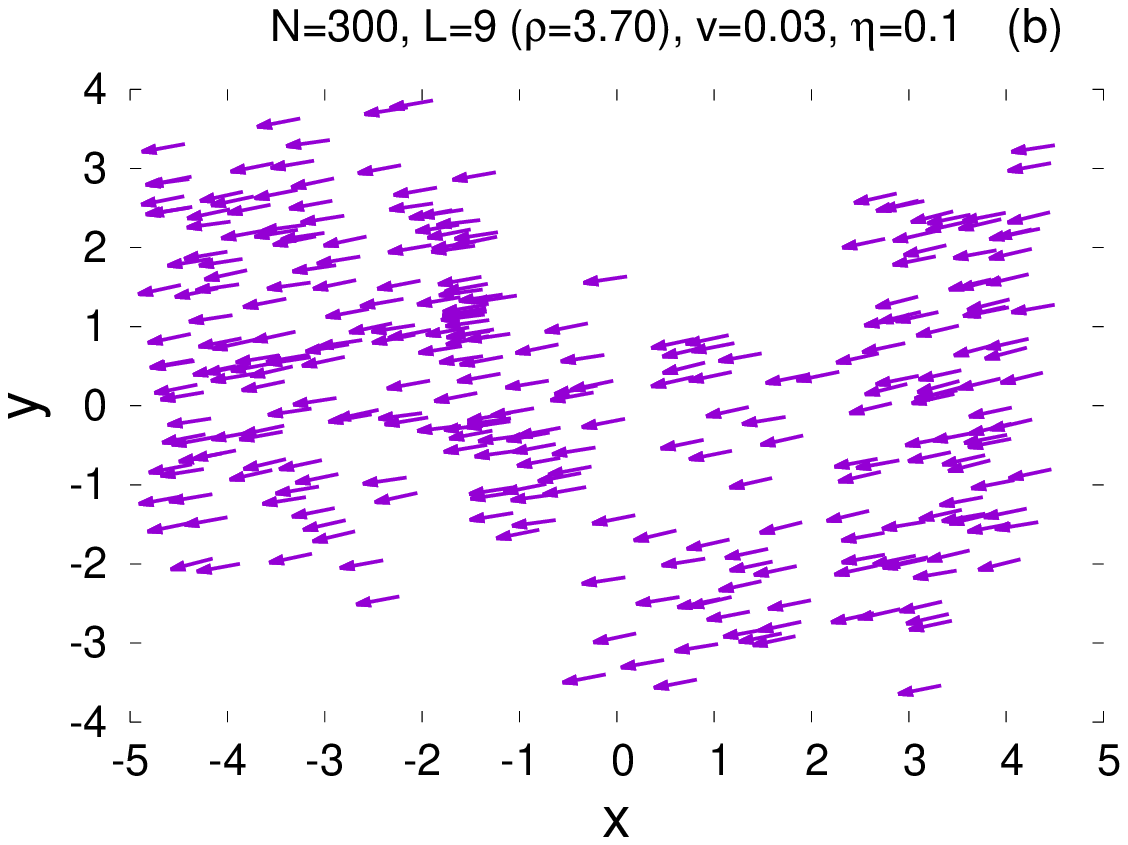}}
\\
\\
\resizebox{7.0cm}{!}{\includegraphics[angle=0]{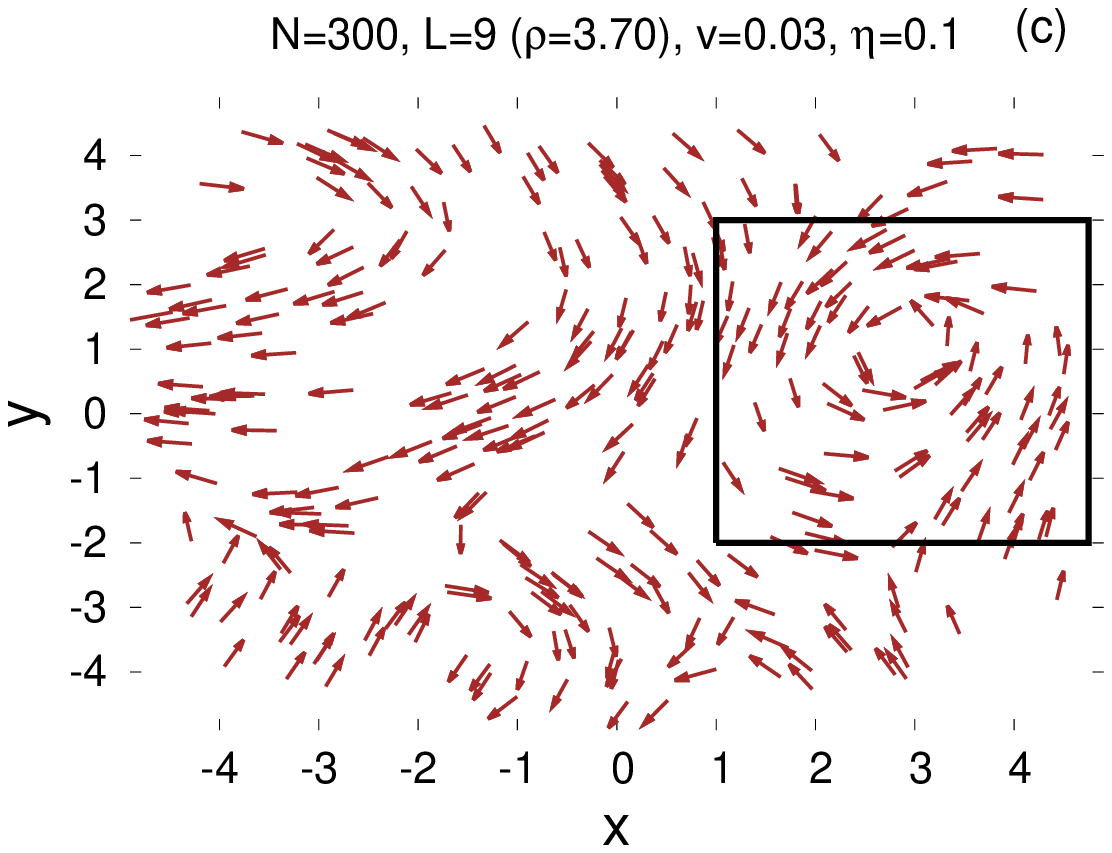}}
\resizebox{7.0cm}{!}{\includegraphics[angle=0]{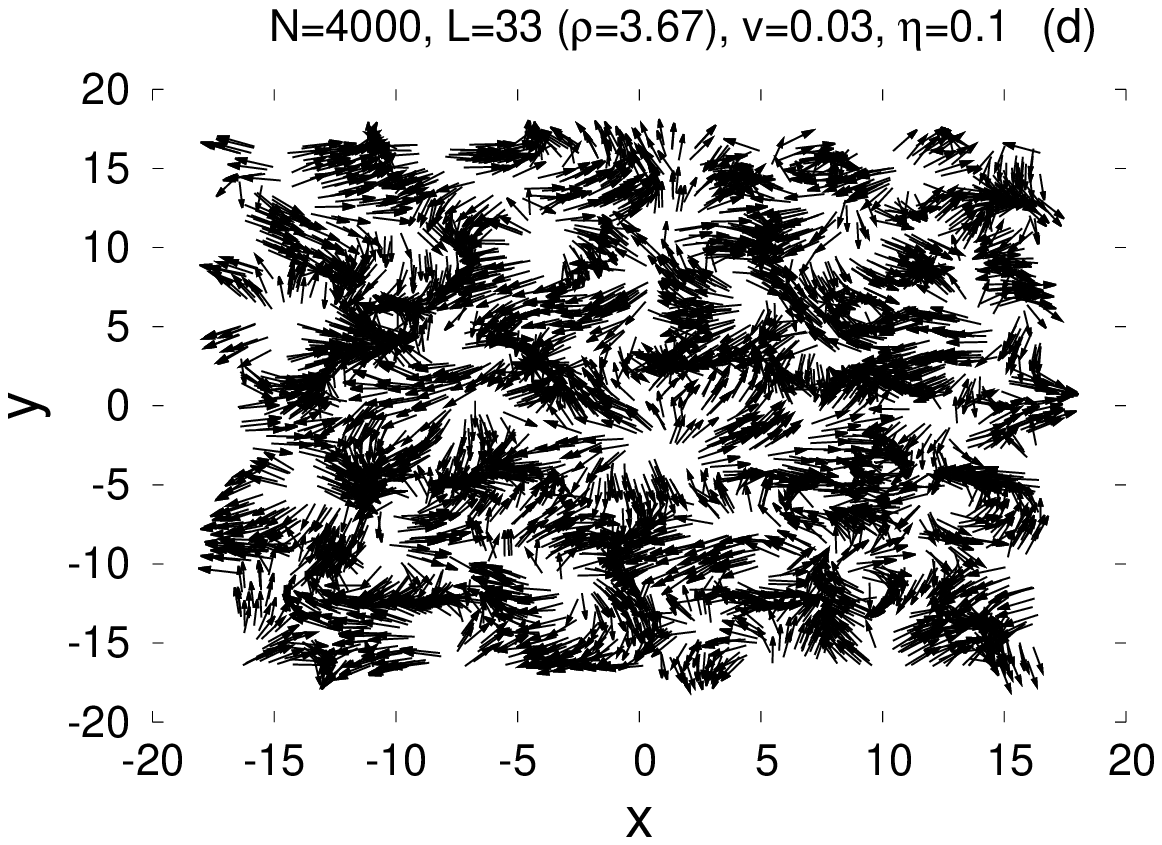}}
\\
\\
\resizebox{7.0cm}{!}{\includegraphics[angle=0]{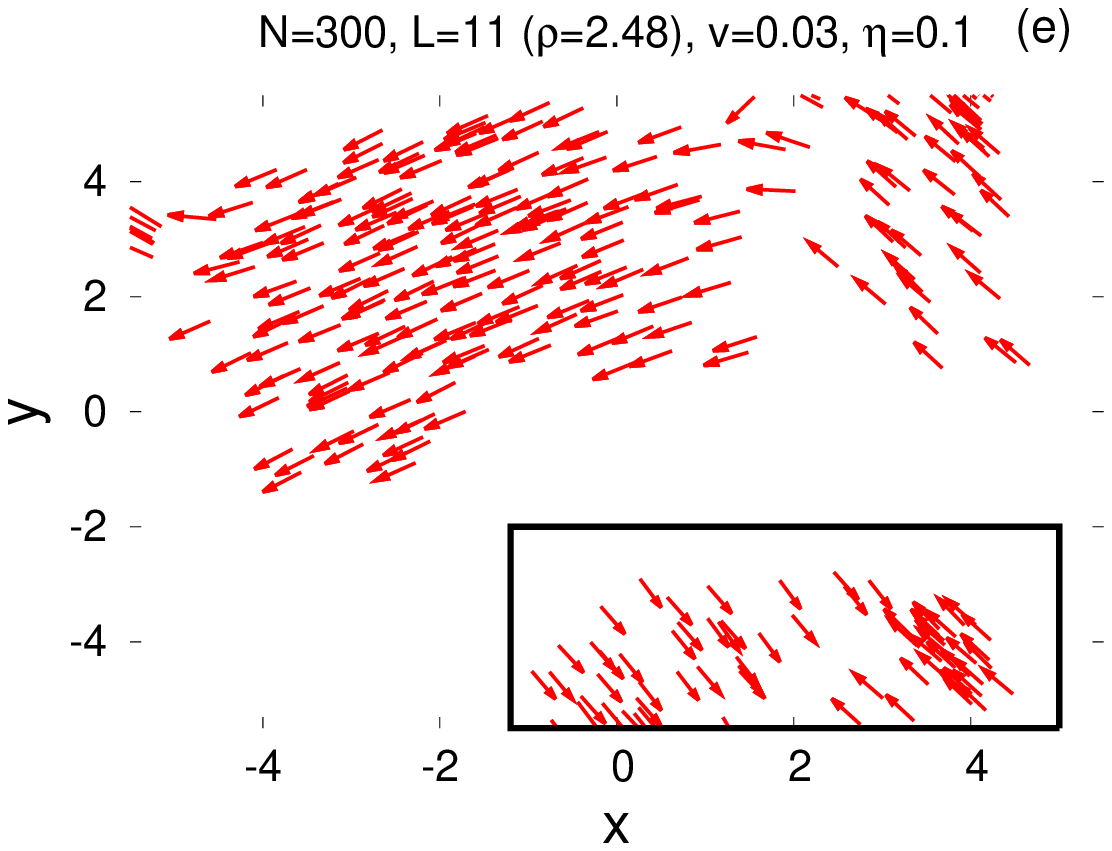}}
\resizebox{7.0cm}{!}{\includegraphics[angle=0]{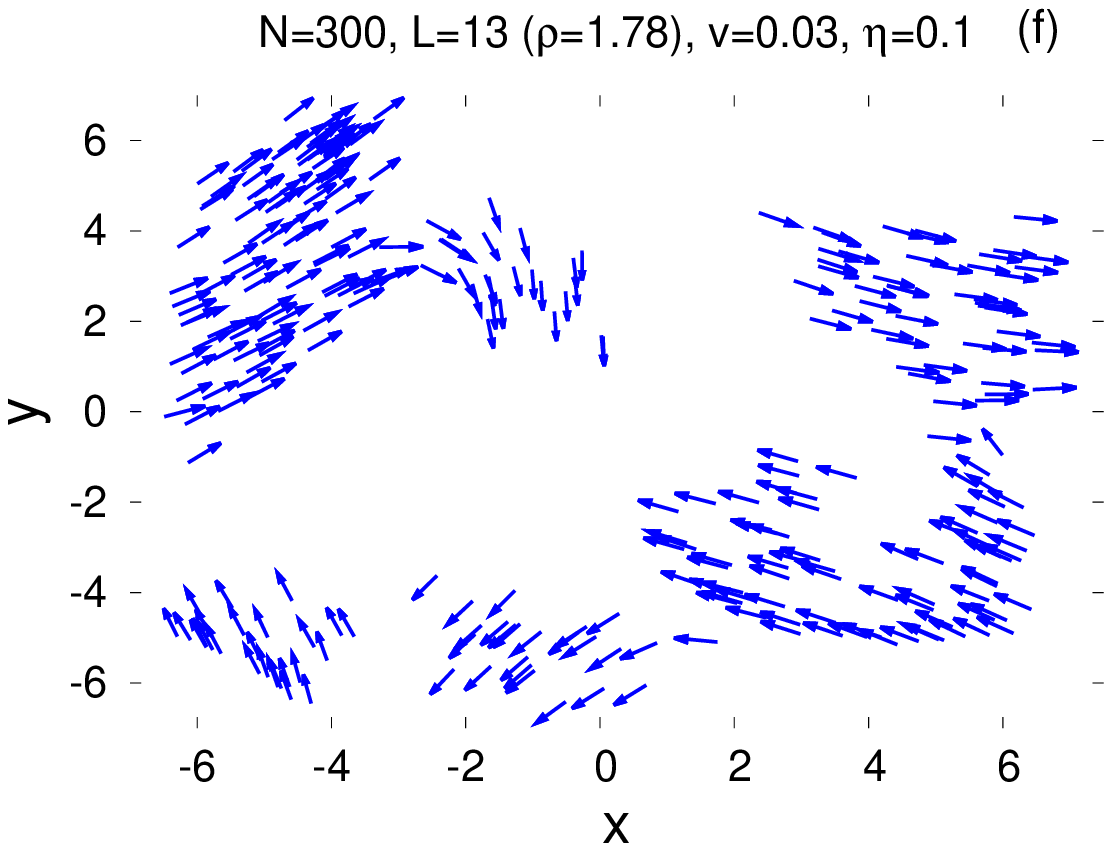}}
          \end{tabular}
\caption{(a) Completely random configuration at $t=0$ for $ N=300, L=9$ $(\rho=3.7)$;
(b) Ordered phase at $t=1000$ at low noise ($\eta=0.1$);
(c) Vortex phase (marked by the black box) at $t=10$ for $ N=300, L=9$ $(\rho=3.7)$;
(d) Number of vortices increases for larger $N(=4000)$ and $L(=33)$ ($\rho=3.67$);
(e) Two groups of oppositely moving particles collide (marked by the black box) 
before changing their directions (at $t=200$, for $N=300, L=11 (\rho=2.48)$);
(f) Multi-domain phases appear at low noise at lower density 
$\rho=1.78 (N=300, L=13)$ at $t=240$.}
\label{fig:morph}
\end{center}
\end{figure}

\begin{figure}[h]
\begin{center}
\begin{tabular}{c}
\resizebox{7.0cm}{!}{\includegraphics[angle=0]{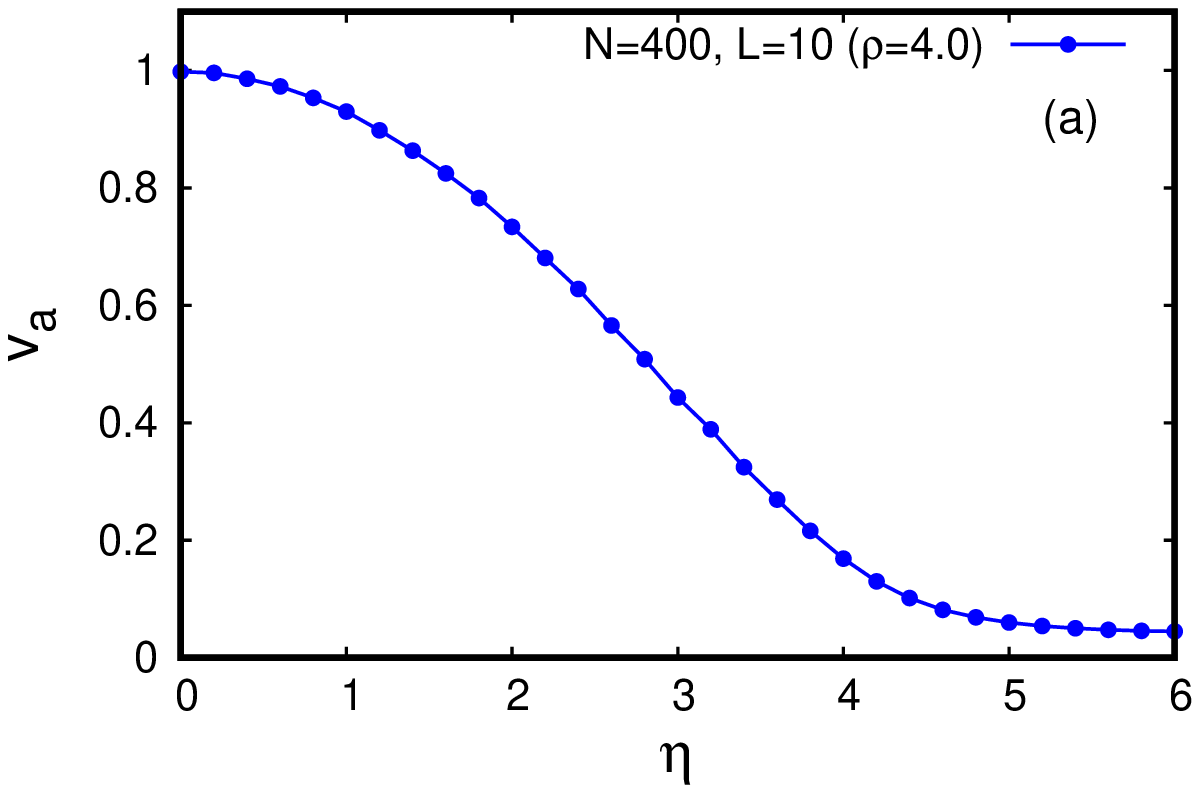}}
\resizebox{7.0cm}{!}{\includegraphics[angle=0]{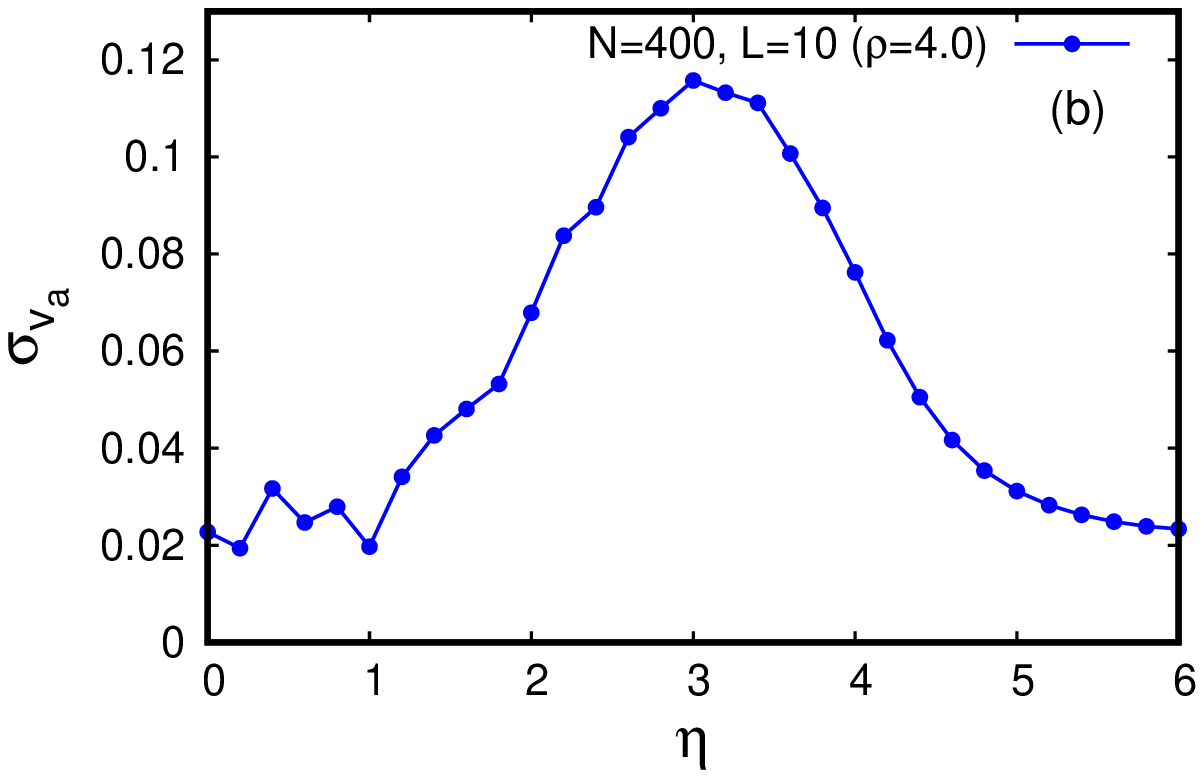}}
\\
\\
\resizebox{7.0cm}{!}{\includegraphics[angle=0]{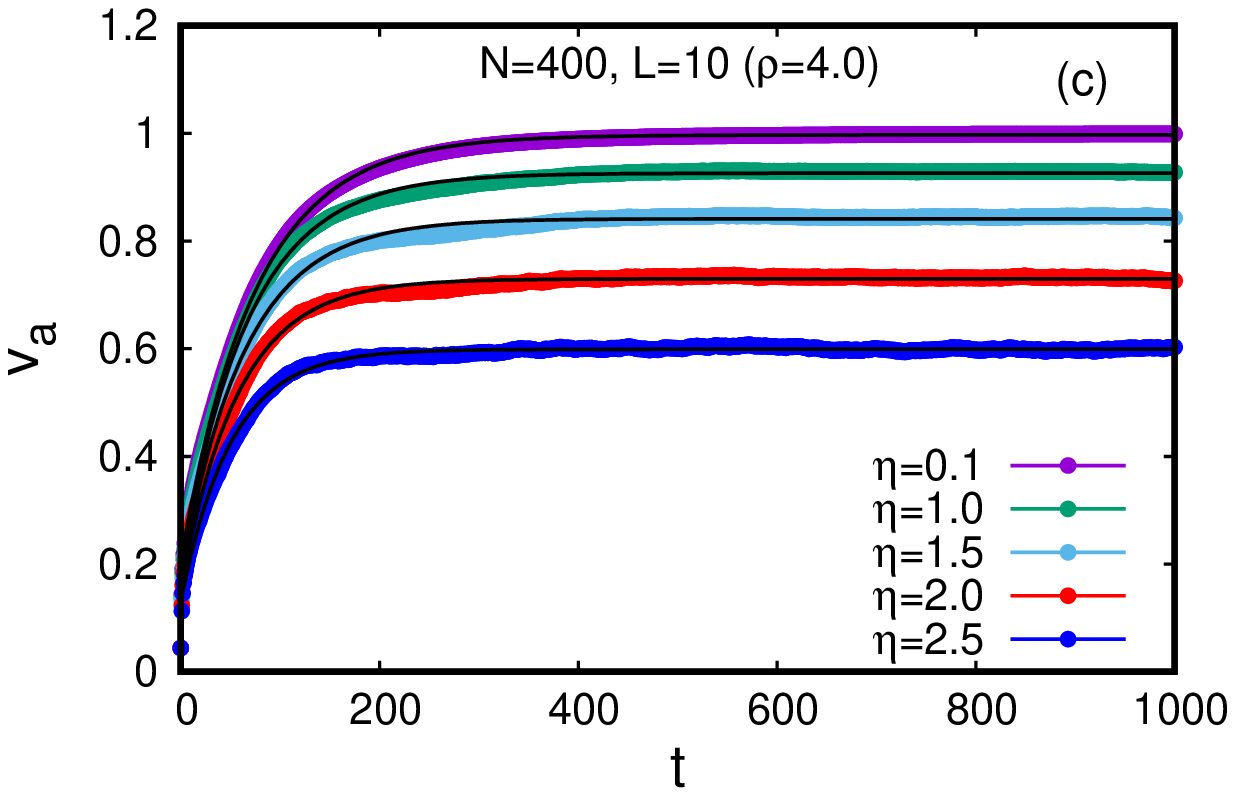}}
\resizebox{7.0cm}{!}{\includegraphics[angle=0]{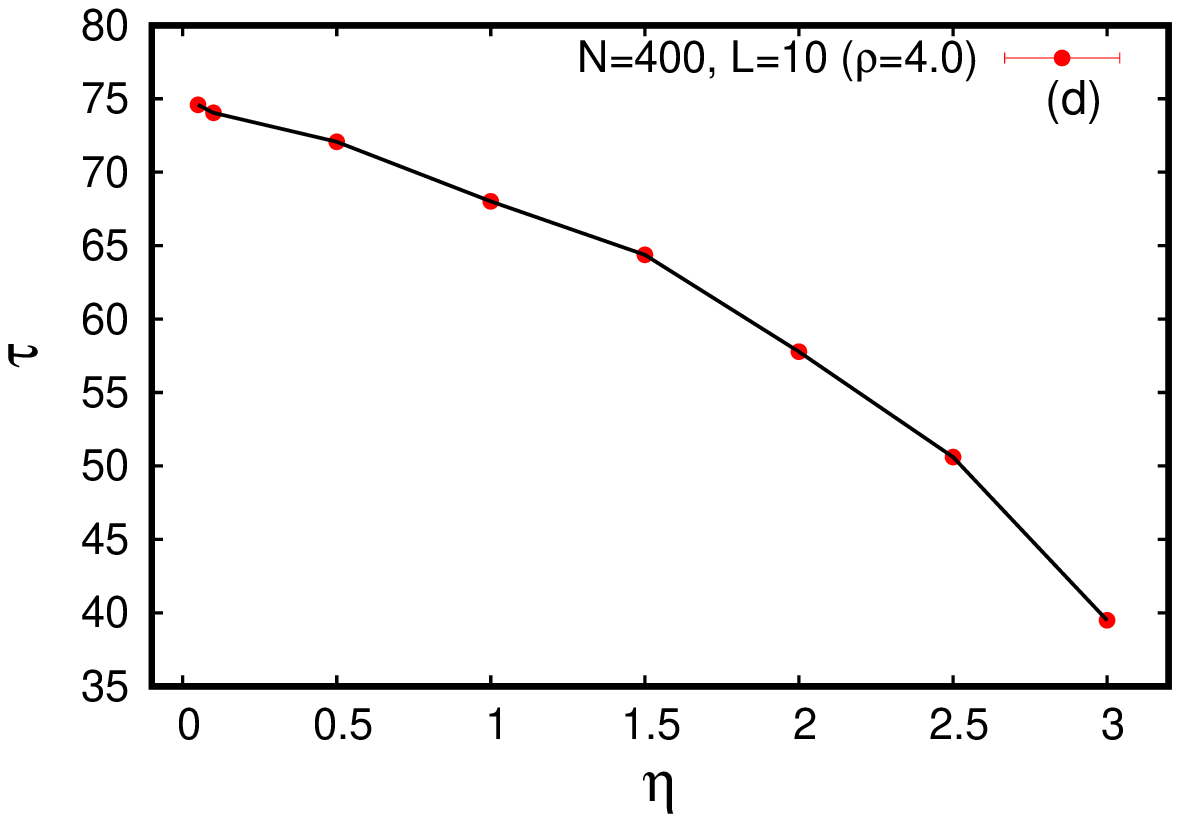}}
          \end{tabular}
\caption{(a) Kinetic phase transition from no transport (order parameter $v_{a}$ very close to $0$ for large $\eta$) to finite net transport ($v_{a}\simeq 1$ for low noise) for $N=400, L=10$;
(b) The critical value of the noise for the phase transition for $N=400, L=10$ is obtained at $\eta_{c}=3.0$ where the fluctuation in the order parameter 
shows maximum;
(c) Time evolution of the order parameter $v_{a}$ is observed for various values of noise which when best-fitted with the equation $ v_{a}=c+d\lbrace 1-exp(-t/\tau)\rbrace $ gives the integrated lifetime $\tau $;
(d) For a fixed density ($\rho=4.0$), the integrated lifetime shows speeding up phenomenon as the noise approaches the critical value.}
\label{fig:noise1}
\end{center}
\end{figure}

\noindent {\bf 3.1. Morphology and characterisation of different transient phases at low noise:}
\vskip 0.1cm
As reported by Vicsek et. al. and reproduced in this study (Fig. \ref{fig:noise1}a), the said system, when subjected to low noise ($\eta=0.1$) well below the critical noise $\eta_{c}$ (Fig. \ref{fig:noise1}b), exhibits spontaneous ordering. It starts from completely random configuration at $t=0$ (Fig. \ref{fig:morph}a) and eventually achieves a global dynamically ordered phase exhibiting long range order in the steady state (at $t= 1000$ in Fig. \ref{fig:morph}b). The time taken to achieve collective behaviour is an intrigue play of noise, velocity and density. However, during the transient time well before achieving the steady state, the system exhibits different types of phenomena depending upon the density of the system. In the Fig. \ref{fig:morph}c, for a system with $N=300$, $L=9$ ($\rho=3.70$),  we have obtained existence of clear vortex which unbinds after a finite lifetime $\tau_{v}$ (vortex region is marked by box in the figure). When both the number of particles and the system size are increased to $N=4000$ and $L=33$, keeping the density more or less same ($\rho=3.67$), the number of vortices is also found to increase (Fig. \ref{fig:morph}d). We have obtained the second type of transient phase when the density was further lowered to $\rho=2.48$ by keeping $N$ fixed and increasing the cell size to $L=11$. Fig. \ref{fig:morph}e depicts that apart from the vortex forming at earlier times, a certain phase exists for a lifetime $\tau_{c}$ where two groups of oppositely moving particles collide and change their directions (collision region is marked by box in the figure). When density is further decreased to $\rho=1.78$ ($N=300$, $L=13$), the third type of transient phase is observed where domains of particles having different directions are formed for a finite lifetime $\tau_{d}$ at different regions
 in the box and possessing different strengths 
(number of particles in a single domain) (Fig. \ref{fig:morph}f).\\

\vskip 0.2cm

\begin{figure}[h]
\begin{center}
\begin{tabular}{c}
\resizebox{7.0cm}{!}{\includegraphics[angle=0]{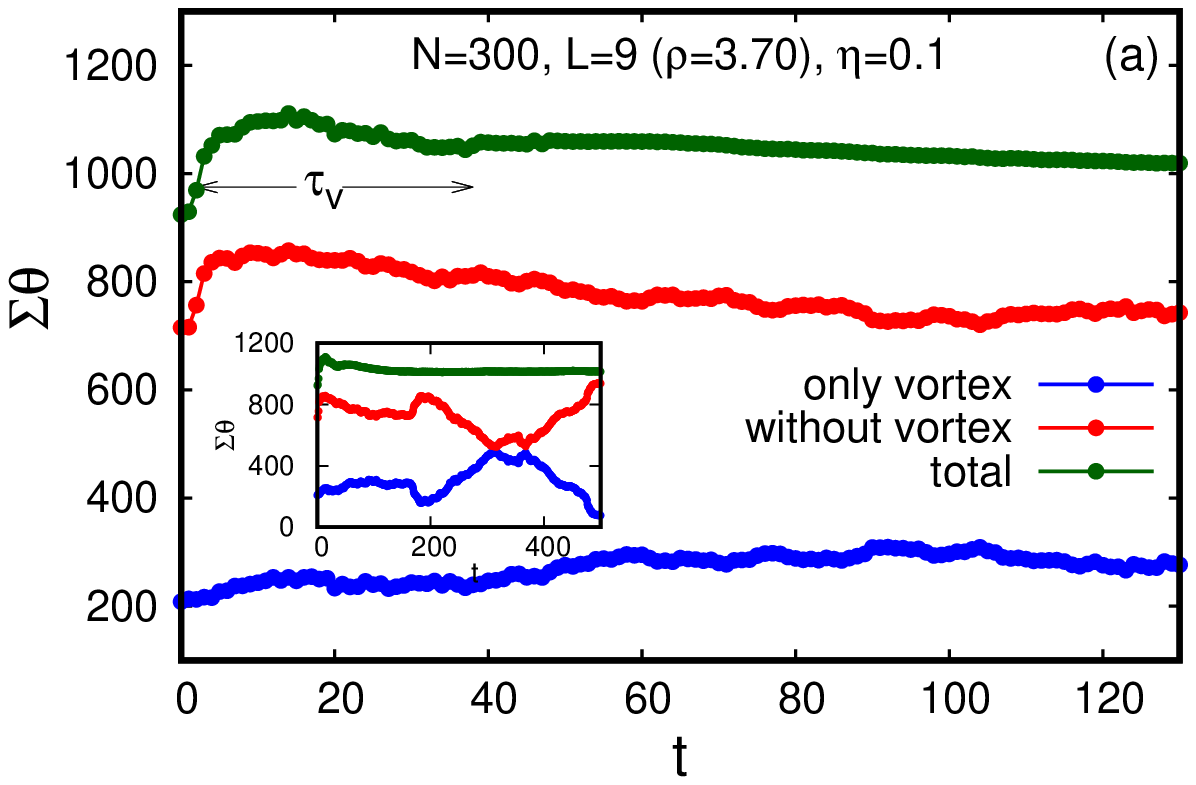}}
\resizebox{7.0cm}{!}{\includegraphics[angle=0]{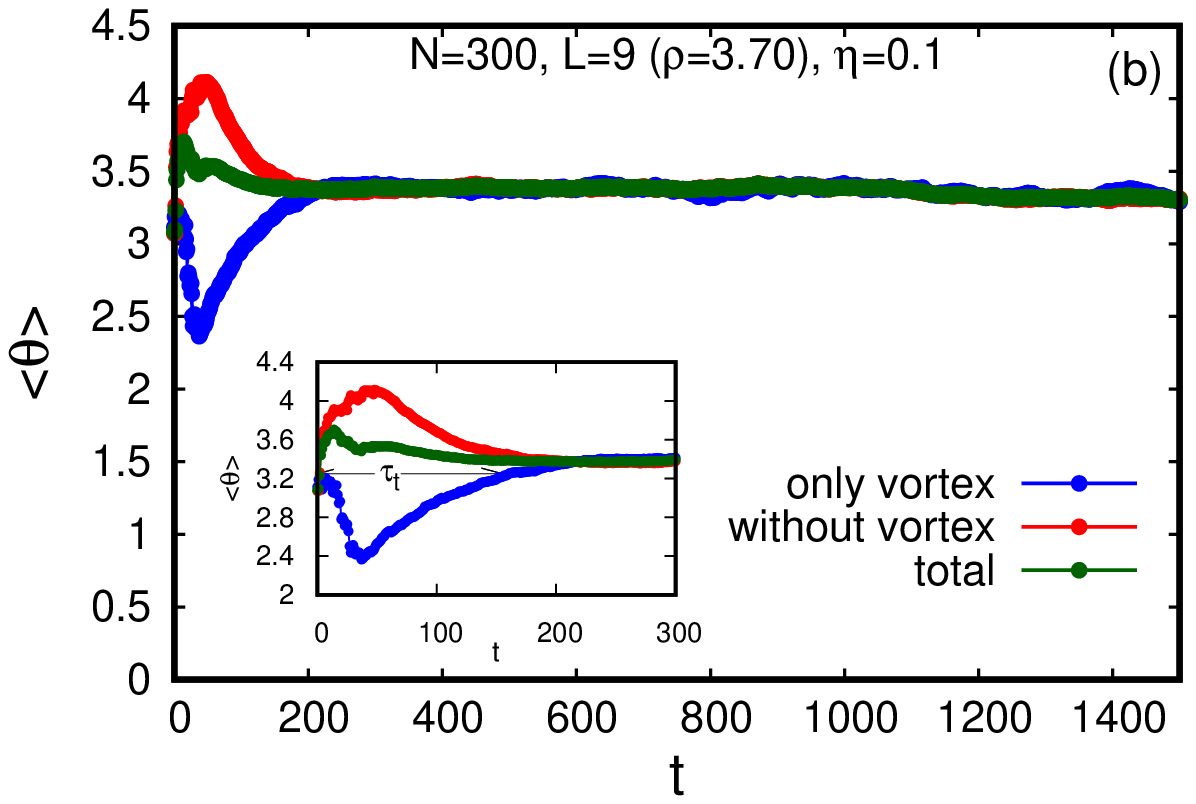}}
\\
\\
\resizebox{7.0cm}{!}{\includegraphics[angle=0]{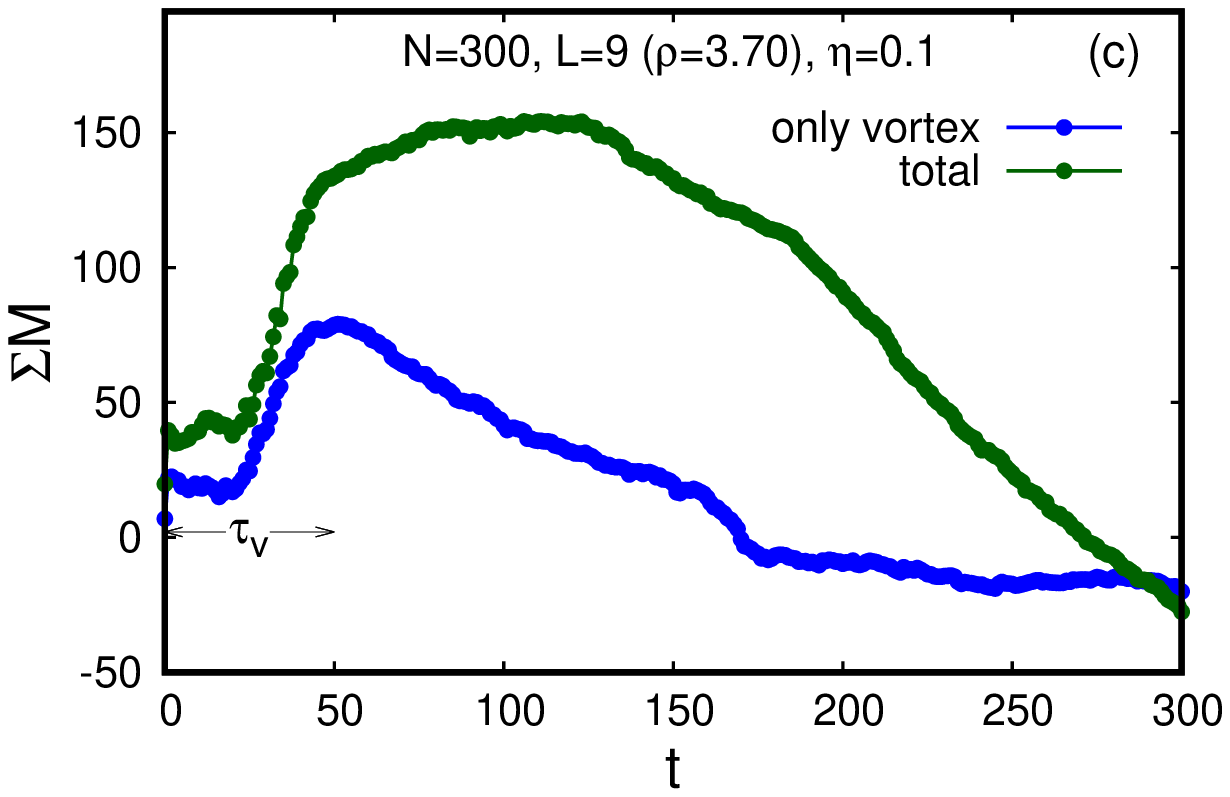}}
\resizebox{7.0cm}{!}{\includegraphics[angle=0]{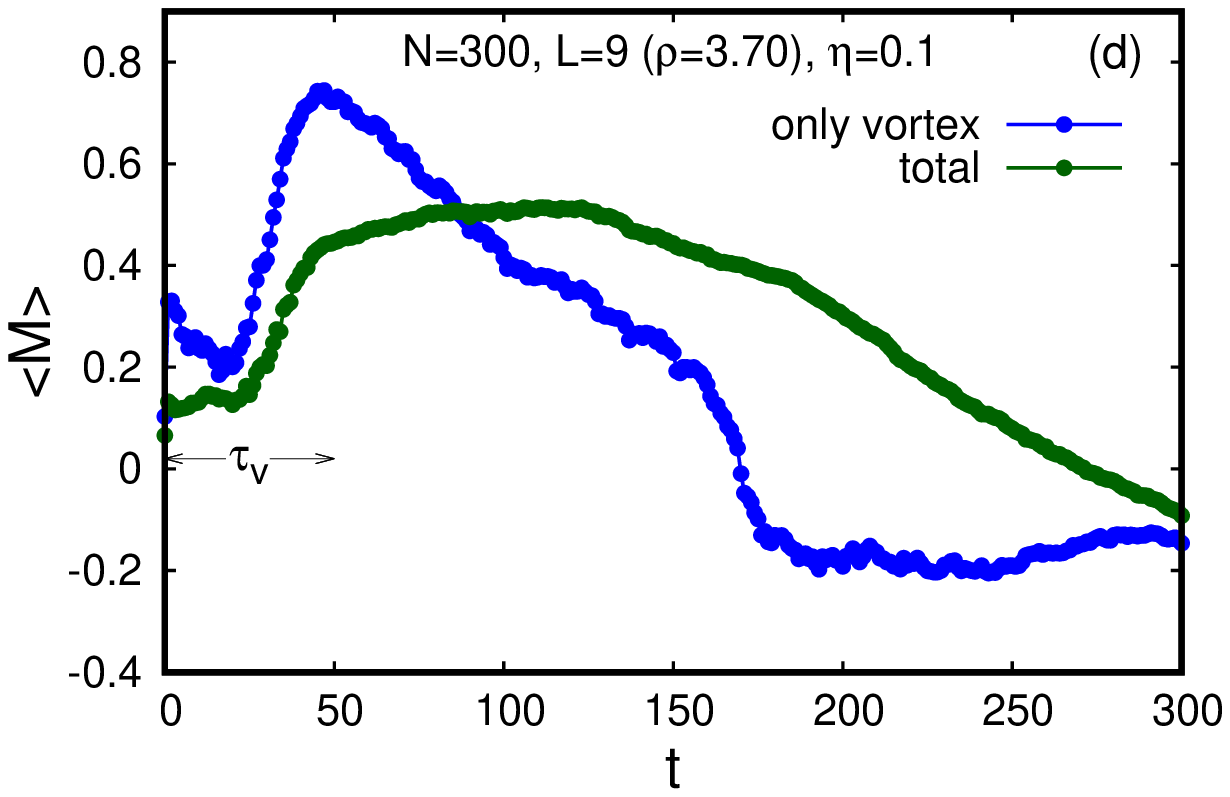}}
          \end{tabular}
\caption{(a) Time evolution of the sum of the angles of particles shows significant contribution by the particles which are whirling inside the vortex (the whole time window is plotted in inset); (b) Contribution by the particles inside the vortex is quite large in the time evolution of the average angle (the whole time window is plotted in inset); (c) Vortex lifetime can be found from the time evolution of the total (unnormalised) angular momentum where the `only-vortex' curve achieves maximum; (d) The same result is verified from the temporal evolution of the average angular momentum which reflects upon the lifetime of the rotating vortex by the maximum in the `only-vortex' curve.}
\label{fig:transient1}
\end{center}
\end{figure}

\noindent {\bf 3.1.1. Rotating Vortex phase:}  
\vskip 0.1cm
In the particular ensemble described in Fig. \ref{fig:morph}c, the vortex is located roughly in the region marked by the black box. The vortex forms immediately after the system starts updating according to the described rule and lives for around 40 discrete time steps. The vortex melts completely after that and a curvature in the velocity portrait morphology exists for sometime before the entire system attains completely directed order (A video of the simulation can be found in YouTube with the link \underline{https://youtu.be/730R0weRacM}).  The contribution due to vortex is measured by separately studying the designated region. We have calculated the time evolution of the sum of the angles of velocity directions of all the particles participating in the vortex and the average of the same (normalised by proper number of particles falling inside the box) and have compared these to the quantities obtained in the complementary cases of particles excluding the vortex region and the total intact system. Fig. \ref{fig:transient1}a shows that the sum of the angles is almost constant for the entire lifetime of the vortex for the `only-vortex' case whereas the complementary `vortex-excluding' part and the total system both show an identical bump in the curve till the vortex exists.  This is due to the fact that, when the vortex rolls out with time, the angles of the particles oppositely directed within the circular vortex cancel out each other and make the sum constant over time. The vortex lifetime $\tau_{v}$ is measured by the width of the bump ($\tau_{v}\approx 40$). The entire time evolution scenario is shown in the inset plot where the `only-vortex' and `vortex-excluding' curves show mirror symmetric behaviour to each other, expectedly, whereas the sum achieves a constant value 
for the total system, since all the particles become uni-directional. The vortex contribution is further verified by calculating the time evolution of normalised average of the angles of directions of the particles (Fig. \ref{fig:transient1}b). The `only-vortex' contribution and the complementary 
`vortex-excluding' contributions show complementary fall and rise in the curves respectively during the transient phase before all of them achieve the same average value denoting collective motion. This deep fall due to the `only-vortex' contribution reflects that the vortices play non-trivial roles in several important studies. Moreover, the lifetime of the entire transient phase ($\tau_{t}$) can be found from Fig. \ref{fig:transient1}b where the normalised average of the angles of directions of the entire system of particles attains steady value. Further studies on the integrated lifetime of the transient phase will again be addressed in Section 3.2.

Another method of finding the vortex lifetime is to measure the temporal evolution of the total angular momentum $\sum M$ (Fig. \ref{fig:transient1}c) and the average angular momentum $\langle M \rangle$ (Fig. \ref{fig:transient1}d) of the particles, where $\sum M = \left| \sum_{i=1}^{N} {\frac{\mathbf{r_{i} \times v_{i}}}{\left| \mathbf{r_{i}} \right| v}} \right|$ and $\langle M \rangle = {1 \over N} \left| \sum_{i=1}^{N} {\frac{\mathbf{r_{i} \times v_{i}}}{\left| \mathbf{r_{i}} \right| v}} \right| $  \cite{calovi}. Both the Fig. \ref{fig:transient1}c and Fig. \ref{fig:transient1}d show that the angular momentum for the `only-vortex' case increases and reaches maximum value at the time up to when the vortex lives and then it starts decreasing as soon as the vortex starts melting. From this, the vortex lifetime can be approximately found to be $\tau_{v} \approx 40$ which is consistent with the previous result. However, for the total system, the angular momentum does increase up to $t \approx 40$, but due to the presence of still-existing curvatures in the velocity portrait of the entire system, it falls at a much slower rate than that for the `only-vortex' case and attains small values when the system eventually achieves steady state ordered phase. This can be explained by the argument that the average angular momentum of the particles tends to achieve higher value when the flock is rotating in a vortex, whereas its values will be less when the particles exhibit unidirectional motion.\\

\vskip 0.2cm

\begin{figure}[h]
\begin{center}
\begin{tabular}{c}
\resizebox{7.0cm}{!}{\includegraphics[angle=0]{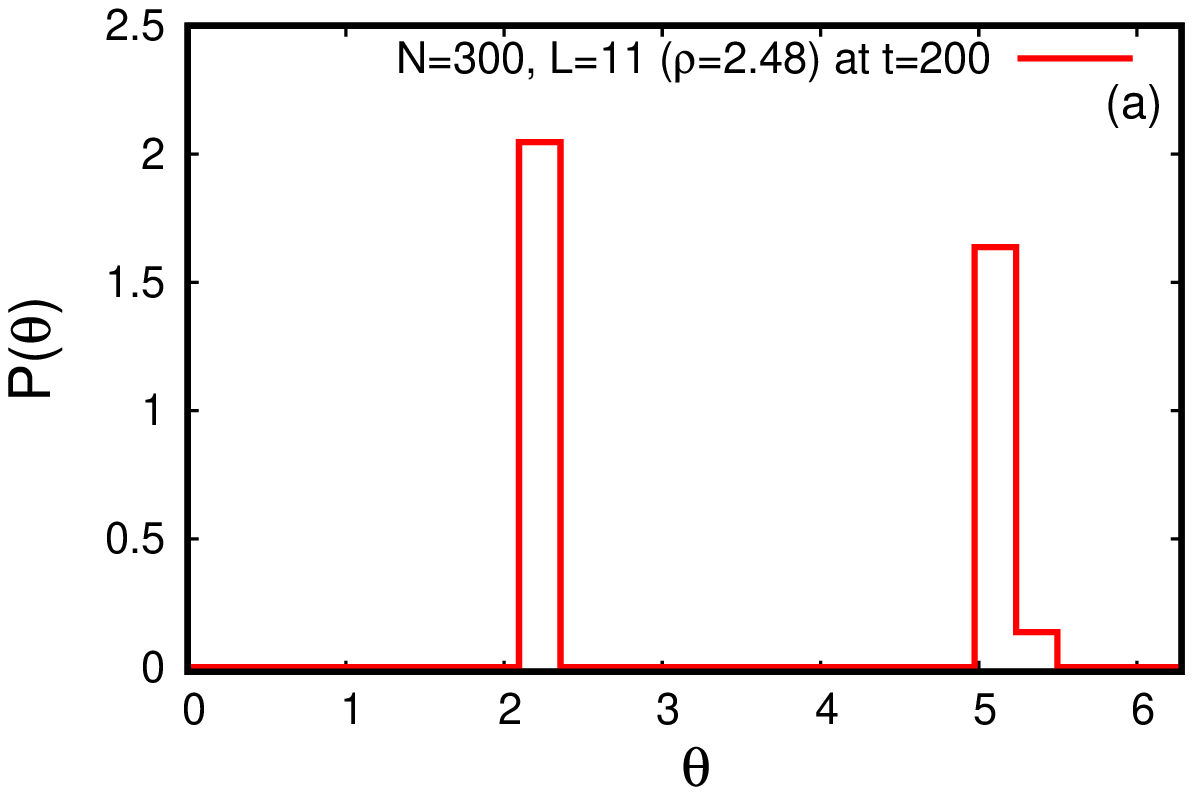}}
\resizebox{7.0cm}{!}{\includegraphics[angle=0]{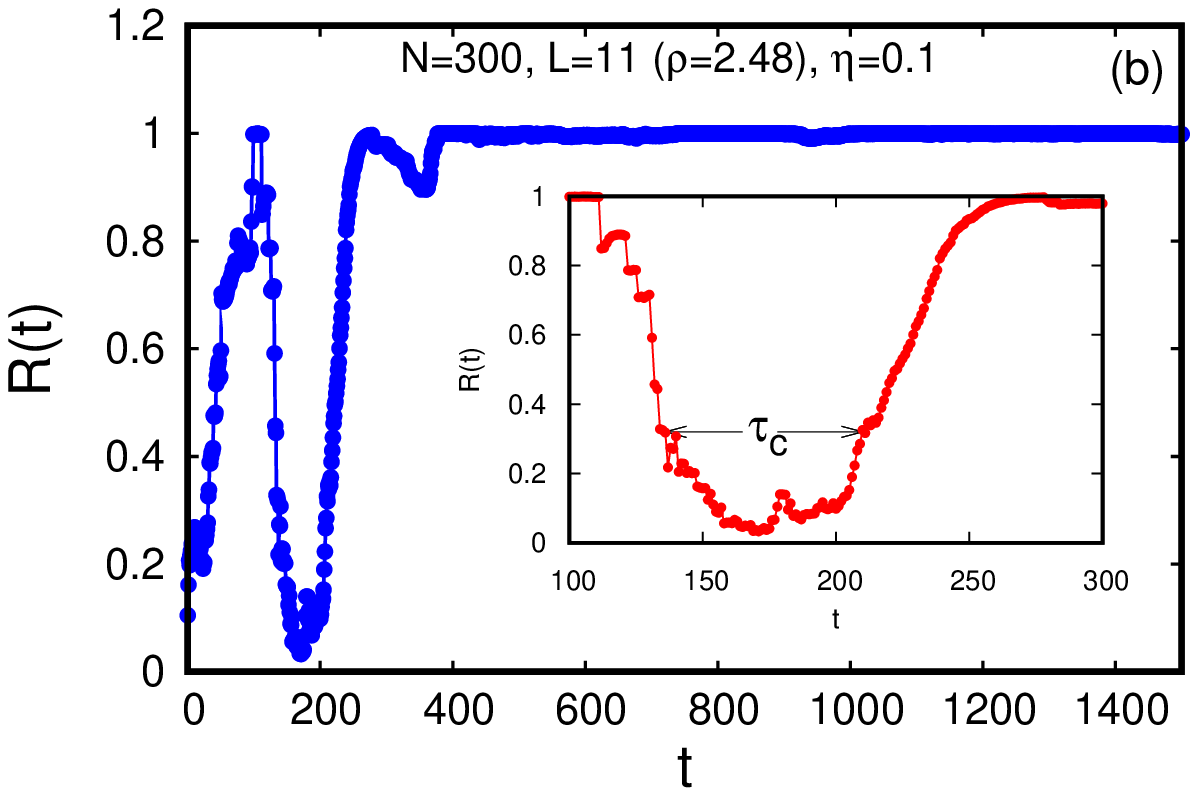}}
\\
\\
\resizebox{7.0cm}{!}{\includegraphics[angle=0]{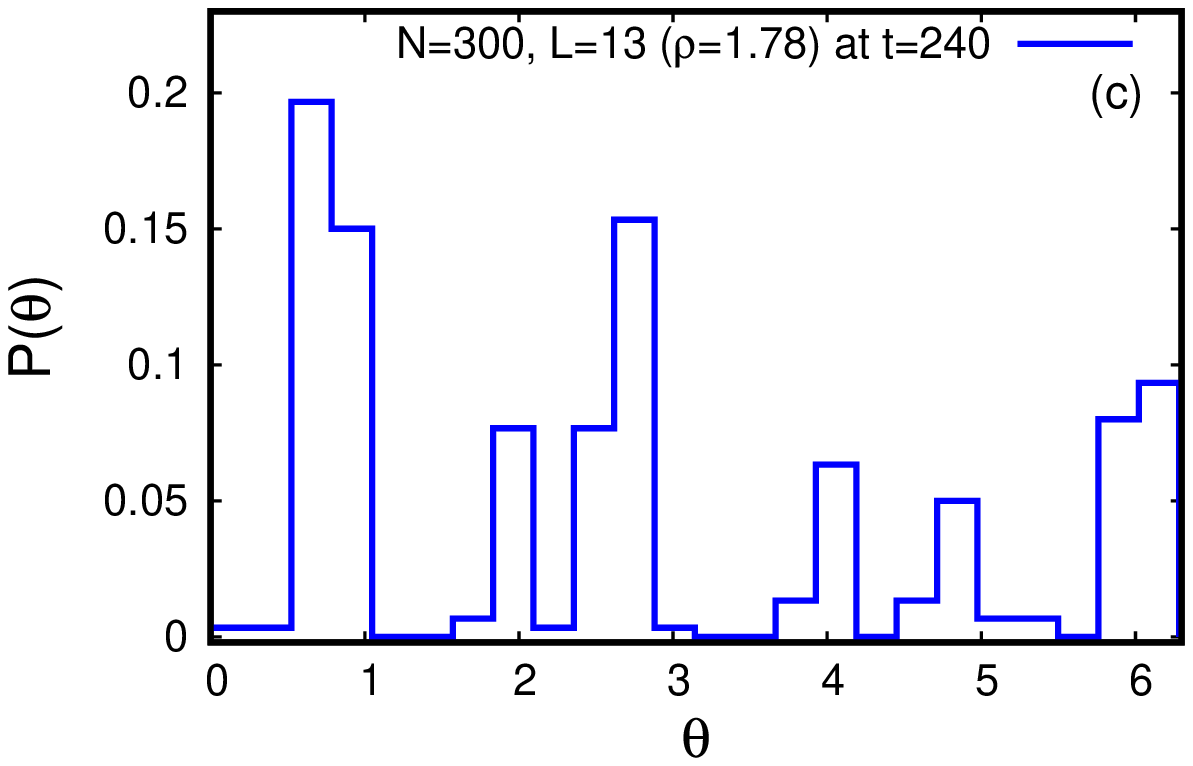}}
\resizebox{7.0cm}{!}{\includegraphics[angle=0]{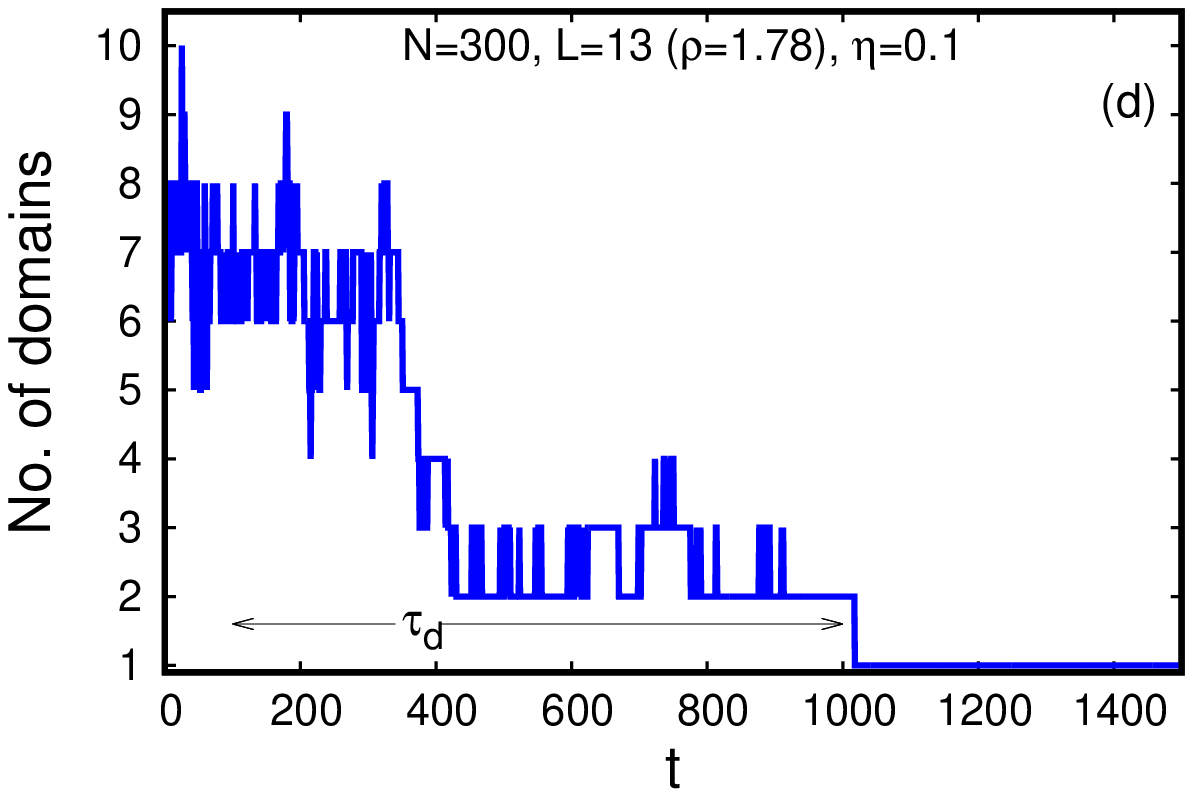}}
          \end{tabular}
\caption{(a) The two distinguished peaks in the normalised angular distribution reflect the collision of two oppositely moving particle groups; (b) The lifetime of the collision phase is marked by the dip in the curve (zoomed-in plot in inset); (c) Six distinguished peaks denote six domains with different strengths and directions; (d) The system starting from initial random configuration gradually forms metastable domains which last upto the time when only one global domain, i.e., ordered phase is reached.}
\label{fig:transient2}
\end{center}
\end{figure}

\noindent {\bf 3.1.2. Colliding phase:}
\vskip 0.1cm
Although decreasing the density lowers the probability of obtaining vortices, it opens door to new kind of transient phases. One such interesting phenomenon observed in the lower density ($\rho=2.48$) regime for the low noise case is collision of particles (A video of the simulation can be found in YouTube with the link \underline{https://youtu.be/uW8TXOZnRs8}) as described in the Fig. \ref{fig:morph}e (colliding particles are marked by black box). This collision phase for this particular ensemble was identified by the normalised probability distribution of the angles of the directions of the particles which shows two distinct peaks corresponding to the two oppositely moving particle groups (Fig. \ref{fig:transient2}a). And these two peaks are approximately separated by $\pi$. The difference in the heights of these two peaks is justified by the strengths of the two groups. We have measured the lifetime of this collision phase by calculating $R(t)=\sqrt{X(t)^{2}+Y(t)^{2}}$ where $X(t)={1 \over n(t)} \sum_{i=1}^{n(t)} cos\theta_{i}(t)$ and $Y(t)={1 \over n(t)} \sum_{i=1}^{n(t)} sin\theta_{i}(t)$; $n(t)$ denotes the number of colliding particles at the time instant $t$. Fig. \ref{fig:transient2}b exhibits the duration of the collision time ($\tau_{c}\approx 75$) by the deep well in the curve (zoomed-in plot in inset). During the collision, the value of $R(t)$ falls (roughly at $t=135$) due to two opposite contributions and then increases again (roughly at $t=210$, i.e., till the collision persists) and reaches steady state value equal to unity as the colliding particles adopt the mutual uni-directional flow.\\

\vskip 0.2cm

\noindent {\bf 3.1.3 Multi-domain phase:}
\vskip 0.1cm
On decreasing the density further to $\rho=1.78$, the third type of transient phase is obtained which is the existence of several domains (Fig. \ref{fig:morph}f) of different directions formed at different regions of the cell possessing different strengths (number of particles in a particular domain). The domains are identified by the normalised probability distribution of angles (velocity directions) of the particles which exhibits 6 different peaks in the Fig. \ref{fig:transient2}c depicting 6 different domains at $t=240$ for this particular ensemble. The difference in the peak heights reflects that the domain strengths are not equal, some domains stochastically housing more particles than others. Fig. \ref{fig:transient2}d denotes the number of domains evolving with time. Starting from a completely random initial configuration and passing through transient domain phases, the number of domains falls to steady value of one as a global long-range domain is eventually formed depicting steady state 
{\it polar} order. Domain lifetime $\tau_{d}$ is defined as the total time 
until which the domains persist before reaching the steady state. A video of the simulation can be found in YouTube with the link \underline{https://youtu.be/SQO0asLAQ10}.\\

\vskip 0.3cm

\noindent {\bf 3.2. Dependence of integrated lifetime on density:}
\vskip 0.1cm
In order to find the dependence of the lifetime of these transient phases on controlling parameters of the system, we steer our attention to the integrated lifetime of all these transient phases instead of treating them individually. 
It is quite difficult to estimate the differentiated 
lifetimes of all these transient phases. 
In this and the consecutive section, all the obtained results and prescribed relationships are tested on configurations possessing $100, 200, 400, 800$ particles ($N$) and $1000$ ensembles of each $N$.

\begin{figure}[h]
\begin{center}
\begin{tabular}{c}
\resizebox{7.0cm}{!}{\includegraphics[angle=0]{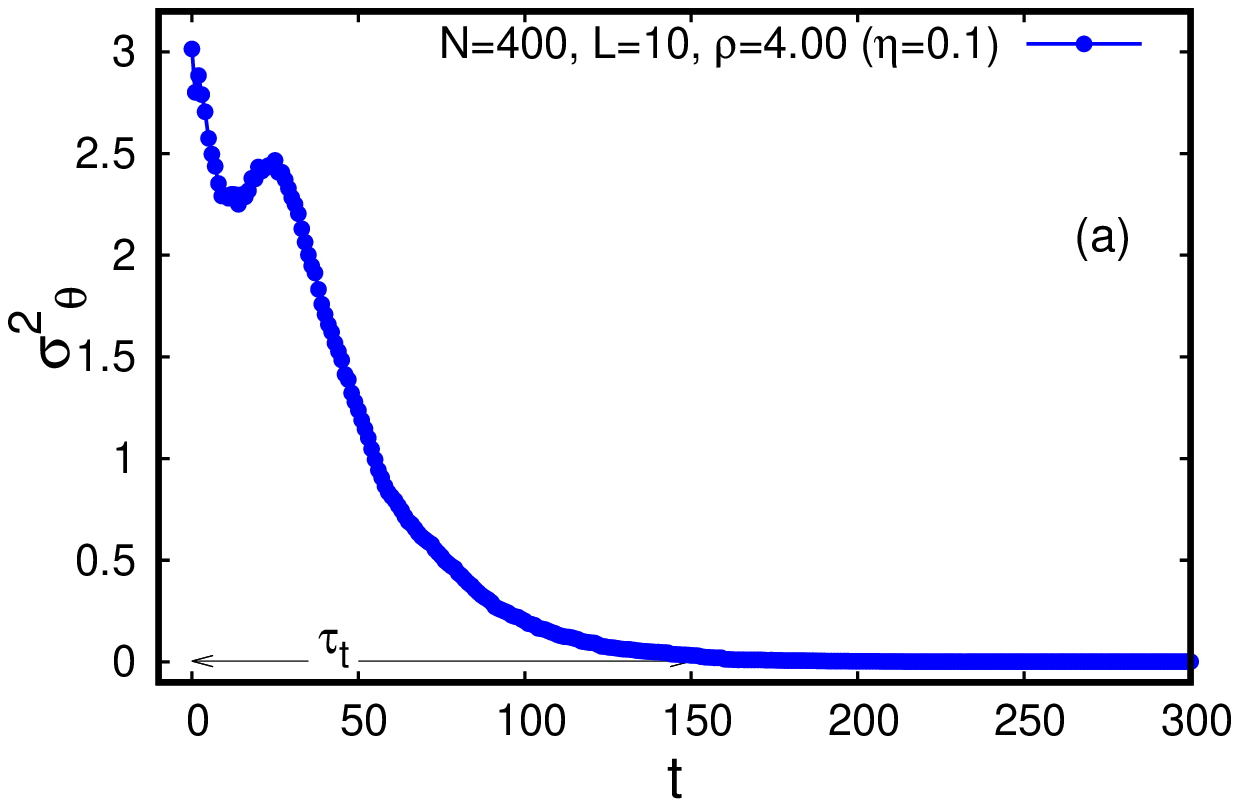}}
\resizebox{7.0cm}{!}{\includegraphics[angle=0]{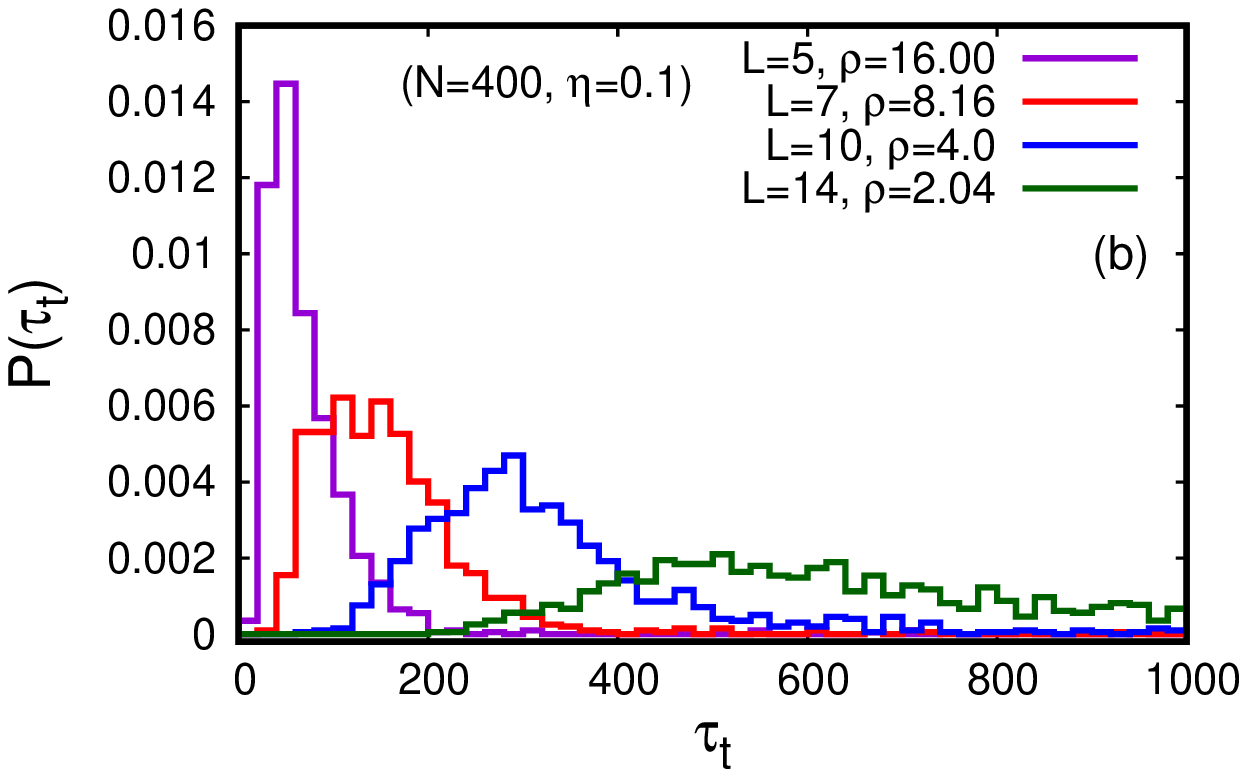}}
\\
\\
\resizebox{7.0cm}{!}{\includegraphics[angle=0]{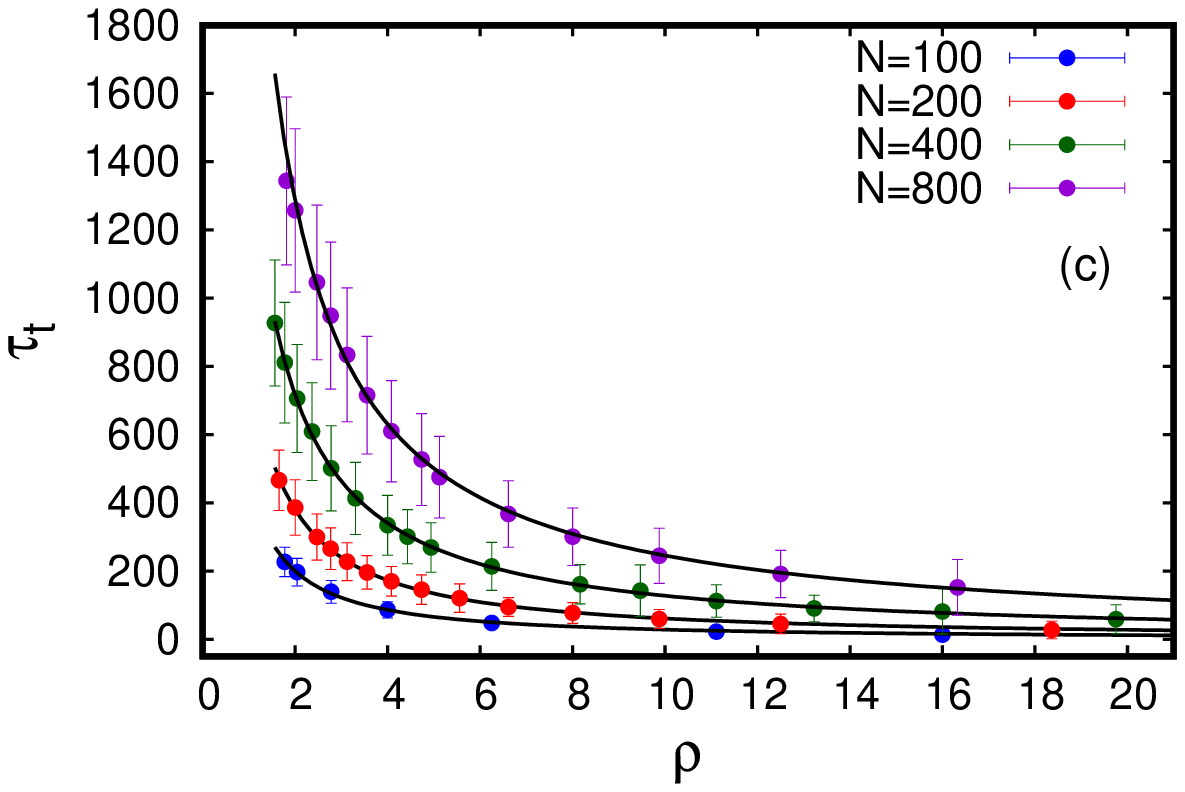}}
\resizebox{7.0cm}{!}{\includegraphics[angle=0]{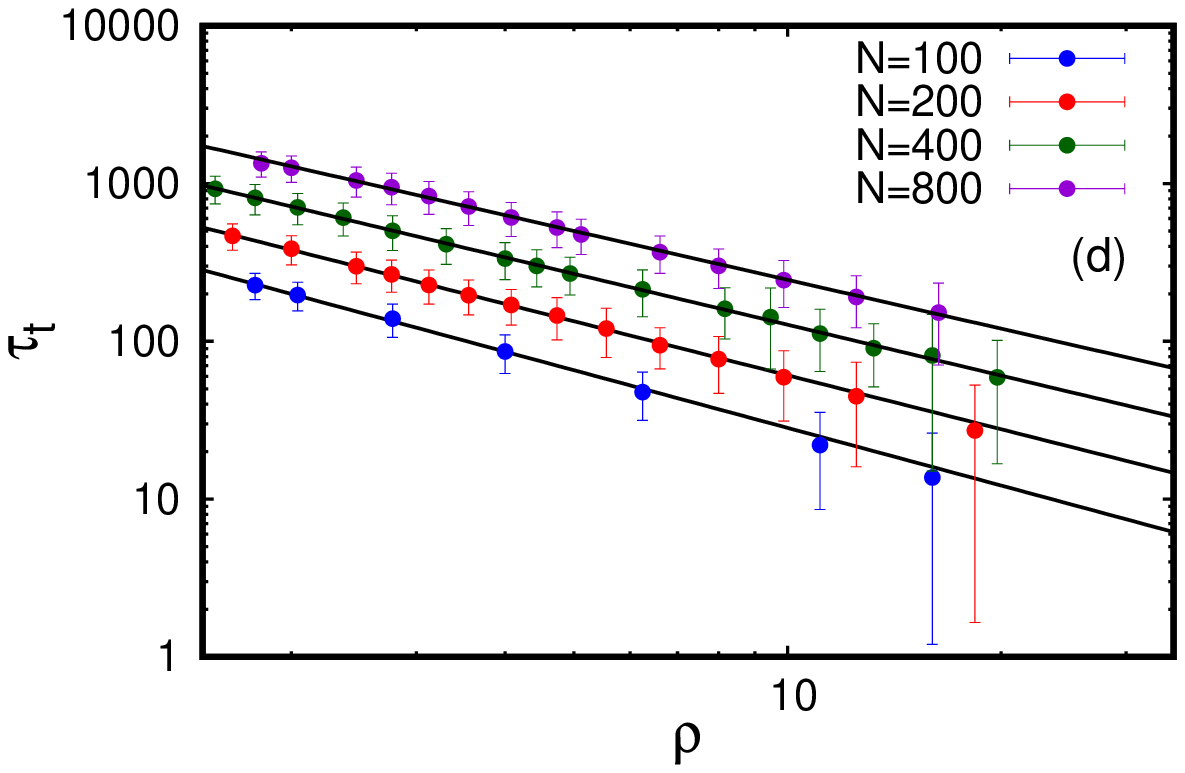}}
          \end{tabular}
\caption{(a) The time evolution of the standard deviation in the angle of direction of the particles for a single ensemble;
(b) The normalised distribution of the lifetime corresponding to 1000 ensembles for different densities;
(c) The power law scaling of dependence of ensemble-averaged lifetime on density (low noise case);
(d) The log-log plot of the same which shows fair straight line (scale invariance nature of the integrated lifetime of transient phases).}
\label{fig:density1}
\end{center}
\end{figure}

The integrated lifetime of the transient phases in the low noise regime ($\eta = 0.1$) is calculated from the time evolution of the standard deviation of the 
distributed angles of directions of the particles $\sigma ^{2}_{\theta} = \langle \theta ^{2} \rangle - \langle \theta \rangle ^{2}$. This $\sigma ^{2}_{\theta}$  of all the particles goes to zero (values $ \leqslant 0.004$ in our simulations) when the system achieves steady ordered state; corresponding time is the 
integrated lifetime ($\tau _{t}$) of the total transient phase (Fig. \ref{fig:density1}a, the plot represents a single realisation of a system having $N=400, L=10$ ($\rho=4.0$)). In the figure, there is a knee-like bending in the curve which is in close correspondence with the vortices, but prescribing any definite relationship between the two is beyond the scope of this current study. 

However, this knee-like bending is found to appear stochastically when single realisations are considered. Upon averaging over many ensembles, 
it washes out this knee-like bending, due to continuous symmetry and stochastic nature. Hence, in order to incorporate ensemble-average picture, the distribution of $\tau_{t}$
(for any fixed set of values of density and noise) is found over 1000 different initial configurations. 
Similarly, different distributions are found for different values of densities
(Fig. \ref{fig:density1}b). Hence, the average values of $\tau_t$
are calculated from these normalised distributions, for different densities, in the low noise regime.

\begin{figure}[h]
\begin{center}
\begin{tabular}{c}
\resizebox{7.0cm}{!}{\includegraphics[angle=0]{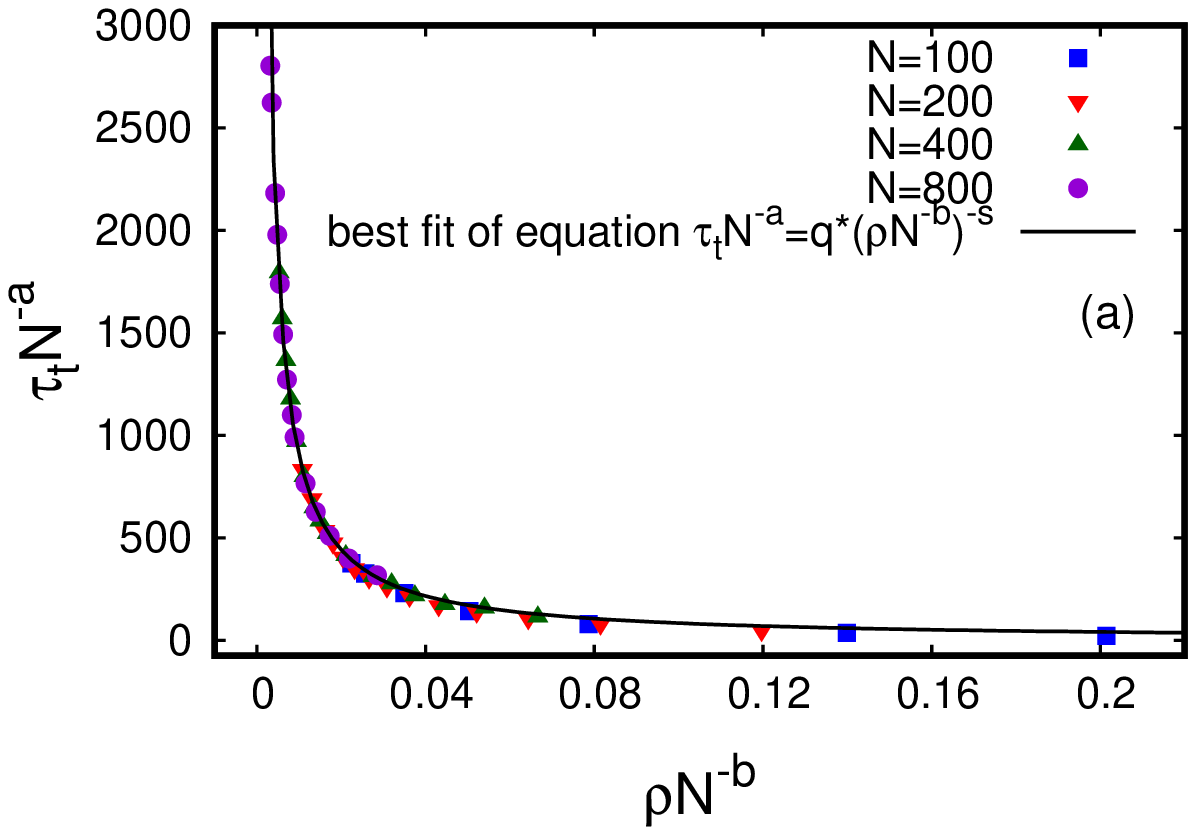}}
\resizebox{7.0cm}{!}{\includegraphics[angle=0]{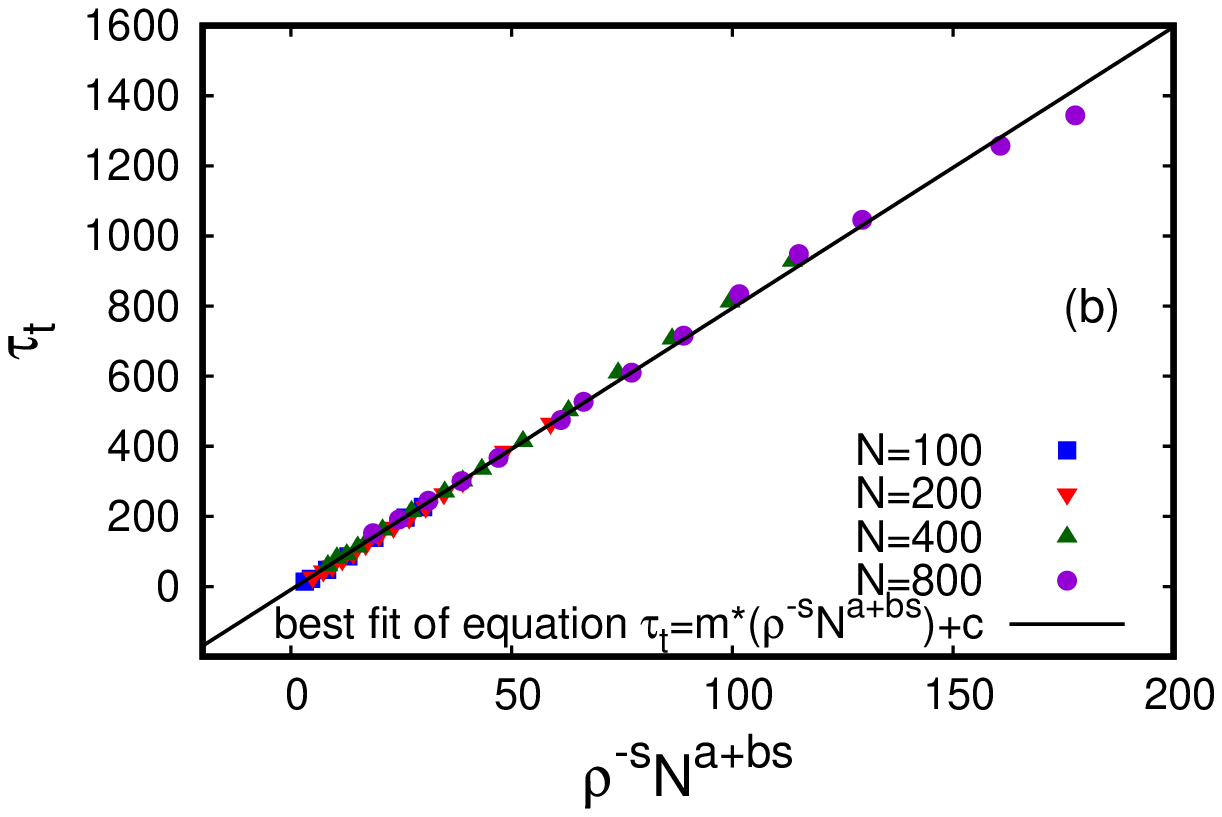}}
\\
\resizebox{7.0cm}{!}{\includegraphics[angle=0]{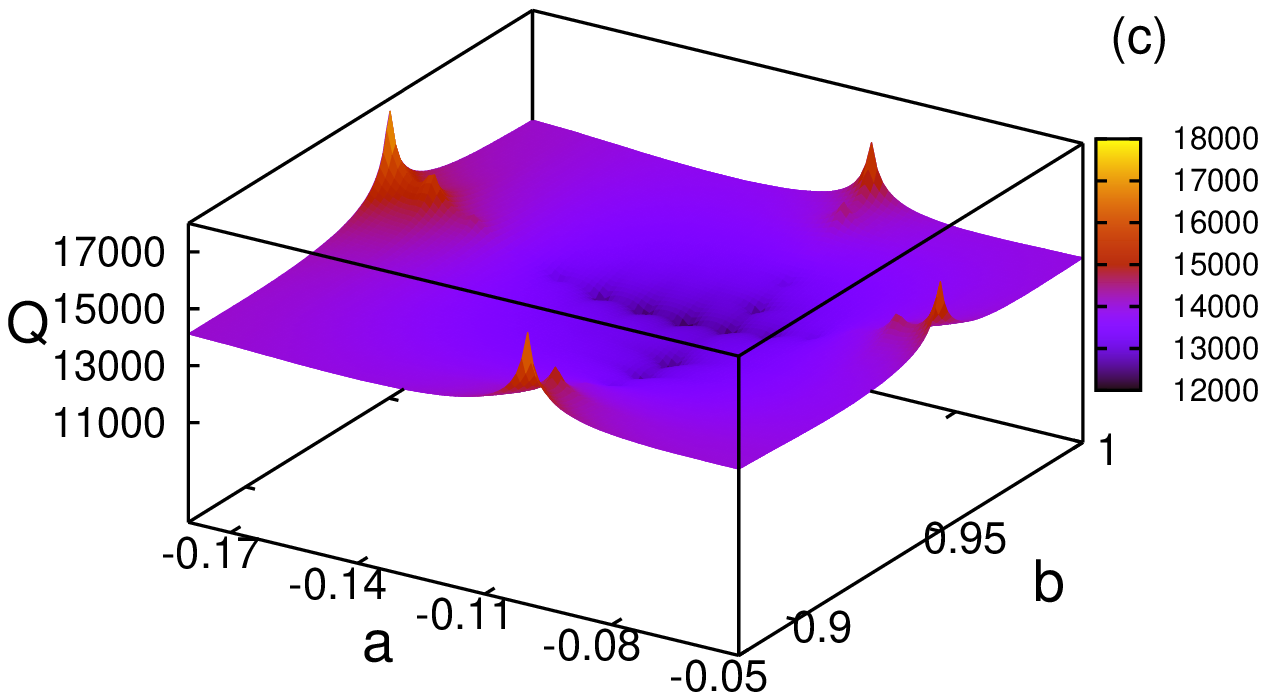}}
\resizebox{7.0cm}{!}{\includegraphics[angle=0]{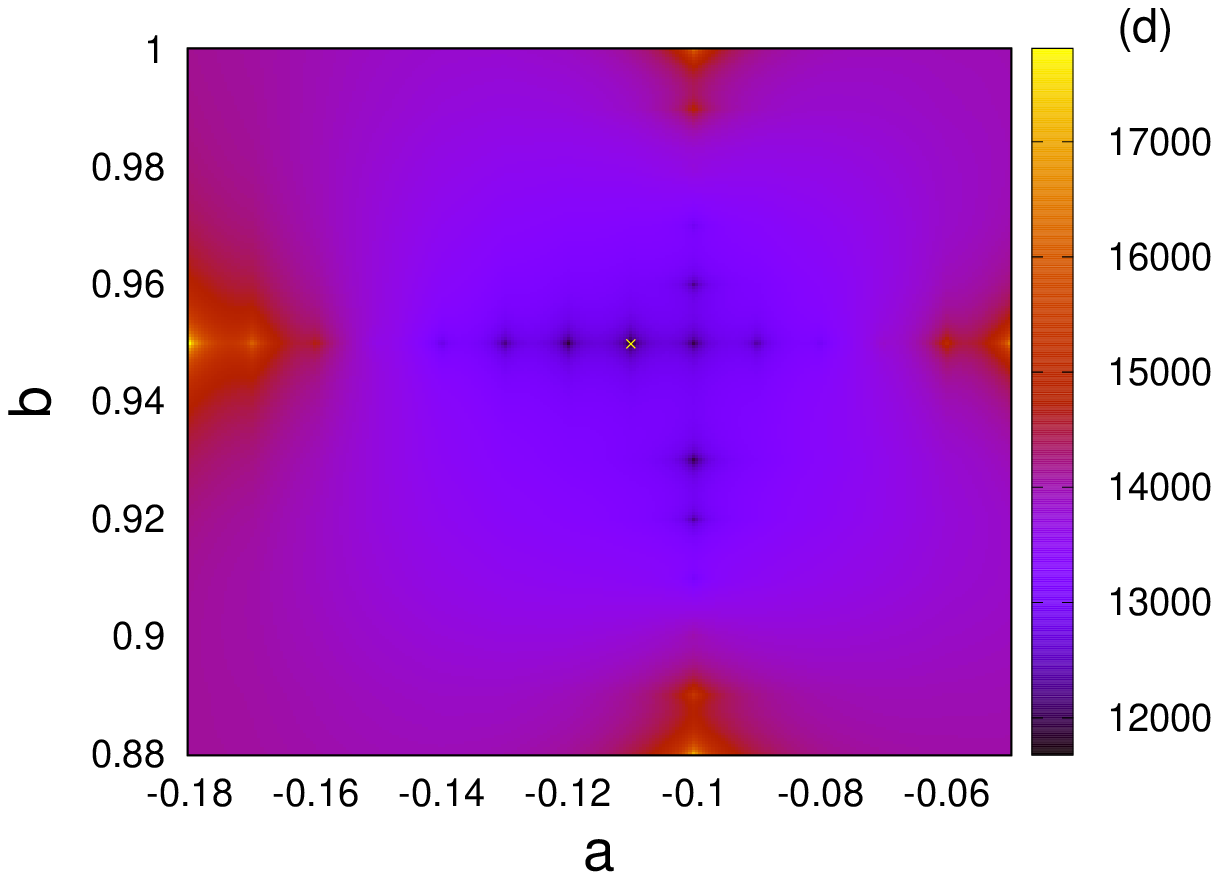}}
\\
\resizebox{7.0cm}{!}{\includegraphics[angle=0]{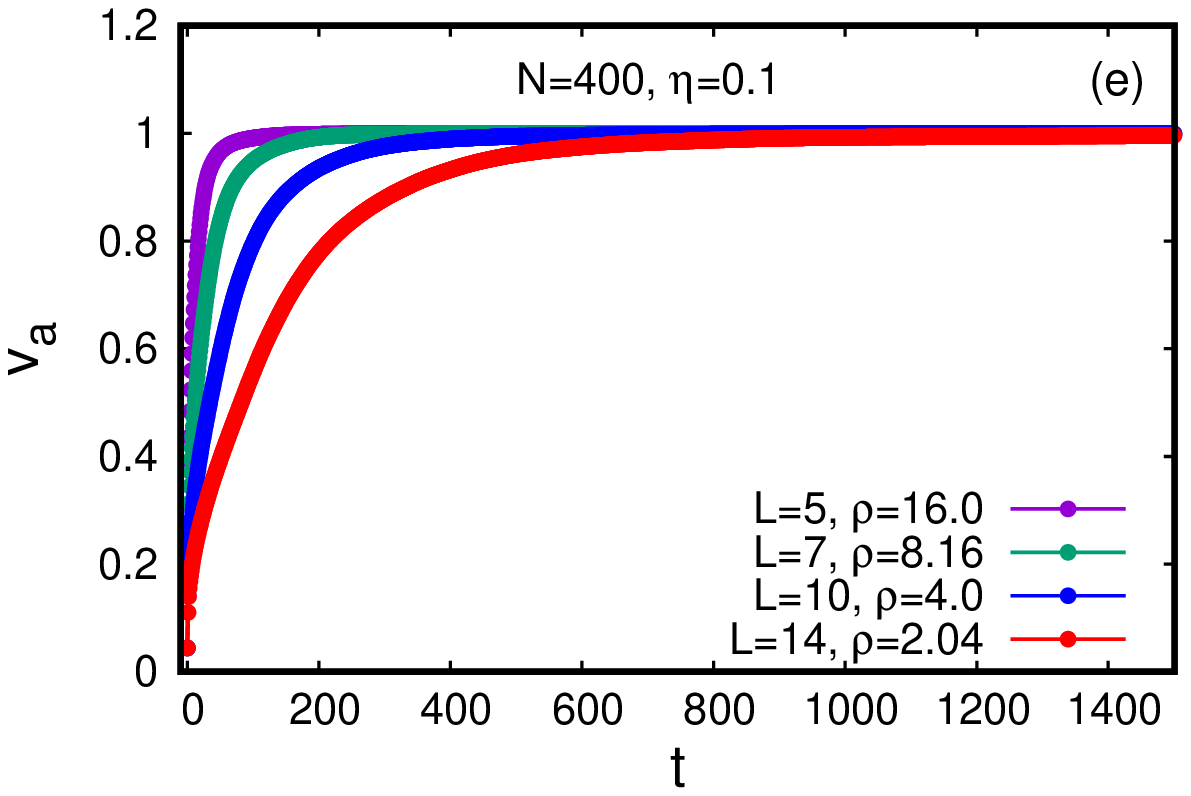}}
          \end{tabular}
\caption{(a) The data-collapsed plot of rescaled integrated lifetime versus rescaled density with the black solid line denoting the best-fit curve giving the power law exponent as $s=1.027\pm 0.008$;
(b) The linear best-fit plot for estimating the data collapsing parameters from the minimum error of the straight line regression;
(c) The best fitting error $Q=\sum_{i}(y_{i}-mx_{i}-c)^{2}$ as a function of the valley of collapse exponents $a,b$;
(d) The global minimum value of the error $Q$($=11581.219$) obtained for the collapse exponents $a=-0.110(\pm0.010)$ and $b=0.950(\pm 0.010)$ which is marked by a cross;
(e) The time evolution of the order parameter $v_{a}$ for different densities (noise fixed at $\eta=0.1$), best-fitting of which by the relation $ v_{a}=c+d\lbrace 1-exp(-t/\tau_{t})\rbrace $ reveals same power law dependence of lifetime on density.}
\label{fig:density2}
\end{center}
\end{figure}

Fig. \ref{fig:density1}c shows the dependence of $\tau_{t}$ on density $\rho$ for $N=100, 200, 400$ and $800$ which reveals a power law scaling. In the log scale, $\tau_{t}$ exhibits fair straight line, i.e., linear dependence on $\rho$ reflecting upon the scale invariance nature of the integrated lifetime (Fig. \ref{fig:density1}d). By the method of data collapse with 
assumed scaling relation, $\tau_{t}N^{-a}=F(\rho N^{-b})$ and best-fitting the curves with $(\tau_{t}N^{-a})=q(\rho N^{-b})^{-s}$ (Fig. \ref{fig:density2}a), the scale invariant power law exponent is obtained to be $s=1.027\pm 0.008$ and the most precise values of the data collapse exponents are obtained from the error minimisation calculations of the best linear fit of $\tau_{t}=m(\rho^{-s} N^{(a+bs)})+c$ 
(Fig. \ref{fig:density2}b). The best fitting error is defined as 
$Q=\sum_{i}(y_{i}-mx_{i}-c)^{2}$,  
where $y_{i}$ and $x_{i}$ represent $\tau_{t}$ and $\rho^{-s} N^{(a+bs)}$ respectively (Fig. \ref{fig:density2}b). 
The surface plot of the error $Q(a,b)$ (taking $a$ and $b$ as axes)
is shown in Fig. \ref{fig:density2}c.   
The global minimum value of $Q$($=11581.219$) is obtained 
for the values of the 
collapse exponents $a=-0.110(\pm0.010)$ and $b=0.950(\pm 0.010)$ 
(marked by a cross in Fig. \ref{fig:density2}d).

The same methodology of finding the dependence of transient lifetime on density does not work for high noise regime since the second moment ($\sigma ^{2}_{\theta}$) achieves saturation at values much larger than 0.004 (which is set as the cut-off value to
represent the ordered phase). Hence, in order to scan the entire range of $\eta$ [$0$,$2\pi$], another methodology is introduced which we have further used rigorously for studying the dependence of the lifetime 
of transient phases on noise when the density is kept fixed (in section 3.3). However, the same is also verified for the already-mentioned low-noise, density-dependent studies of the transient lifetime as well. Fig. \ref{fig:density2}e shows the time evolution of the order parameter $v_{a}$ for different densities (noise fixed at $\eta=0.1$), best-fitting of which by the relation $ v_{a}=c+d\lbrace 1-exp(-t/\tau_{t})\rbrace $ reveals same power law dependence of lifetime on density reported in this section.\\

\vskip 0.3cm

\noindent {\bf 3.3. Dependence of integrated lifetime on noise:}
\vskip 0.1cm

For studying the dependence of integrated lifetime
of the transient phases on noise, density is kept fixed at $\rho=4.0$ and the time evolution of the order parameter $v_{a}$ is observed for various values of noise (Fig. \ref{fig:noise1}c shows a representative system with $N=400$ and $L=10$). The saturation of the order parameter denotes the non-equilibrium steady state condition. Note that the steady state values of the order parameter are in accordance with the phase transition curve shown in Fig. \ref{fig:noise1}a. The critical value of noise here is obtained from Fig. \ref{fig:noise1}b in which the fluctuation in the order parameter ($\sigma_{v_{a}}$) attains maximum
value (which is believed to diverge eventually in the thermodynamic limit)
 at $\eta_{c}(L=10)=3.0$ for this representative case. These curves in Fig. \ref{fig:noise1}c are best fitted according to the relation $ v_{a}=c+d\lbrace 1-exp(-t/\tau)\rbrace $ where $c$ and $d$ are fitting parameters and $\tau$ gives the integrated lifetime of the transient phases, obtained from the time evolution of the order parameter. These best fitted $\tau$'s, when plotted against noise $\eta$, present a remarkable result. It shows that the integrated lifetime of the transient phases decreases with higher values of noise as the noise approaches the critical value (Fig. \ref{fig:noise1}d), which is contrary to the usual case of critical slowing down, observed in the equilibrium kind of phase transitions. In the figure, only the representative $N=400$ case is shown while this {\it reverse} critical slowing down phenomenon is similarly obtained for other values of $N$ as well mentioned in the previous section.

\begin{figure}[h]
\begin{center}
\begin{tabular}{c}
\resizebox{7.0cm}{!}{\includegraphics[angle=0]{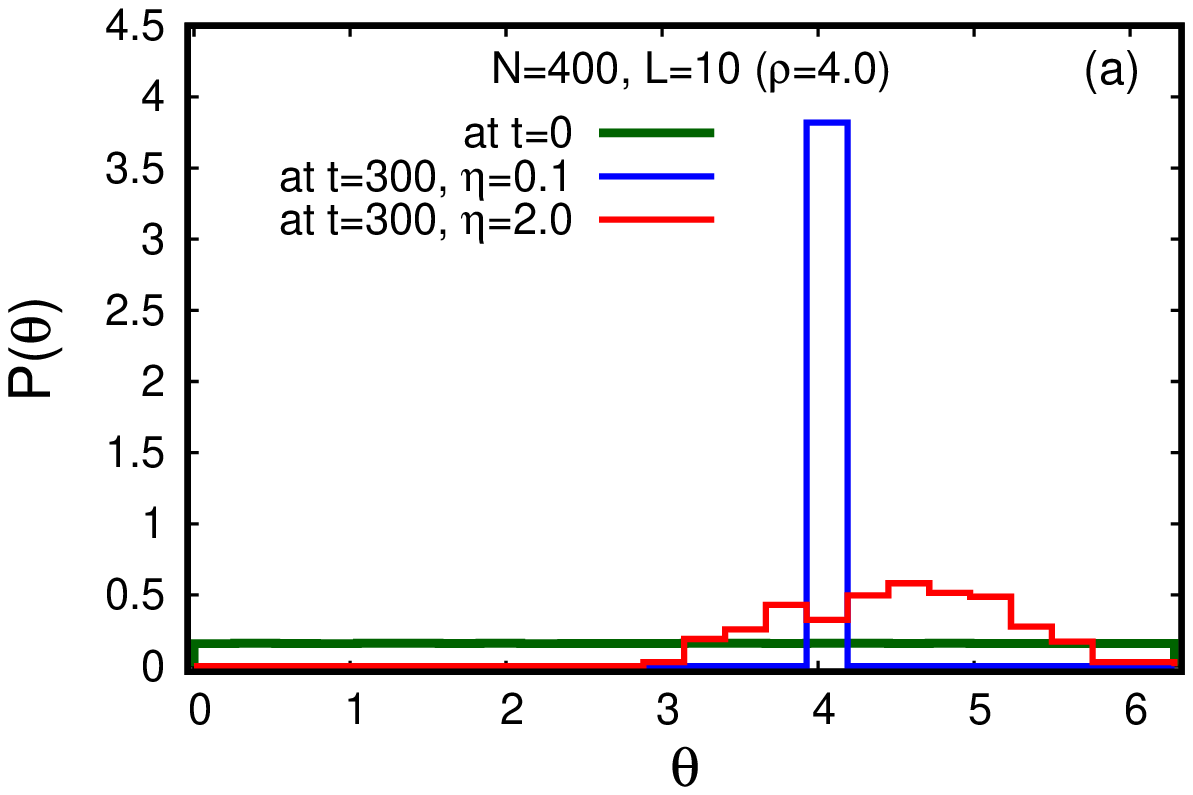}}
\resizebox{7.0cm}{!}{\includegraphics[angle=0]{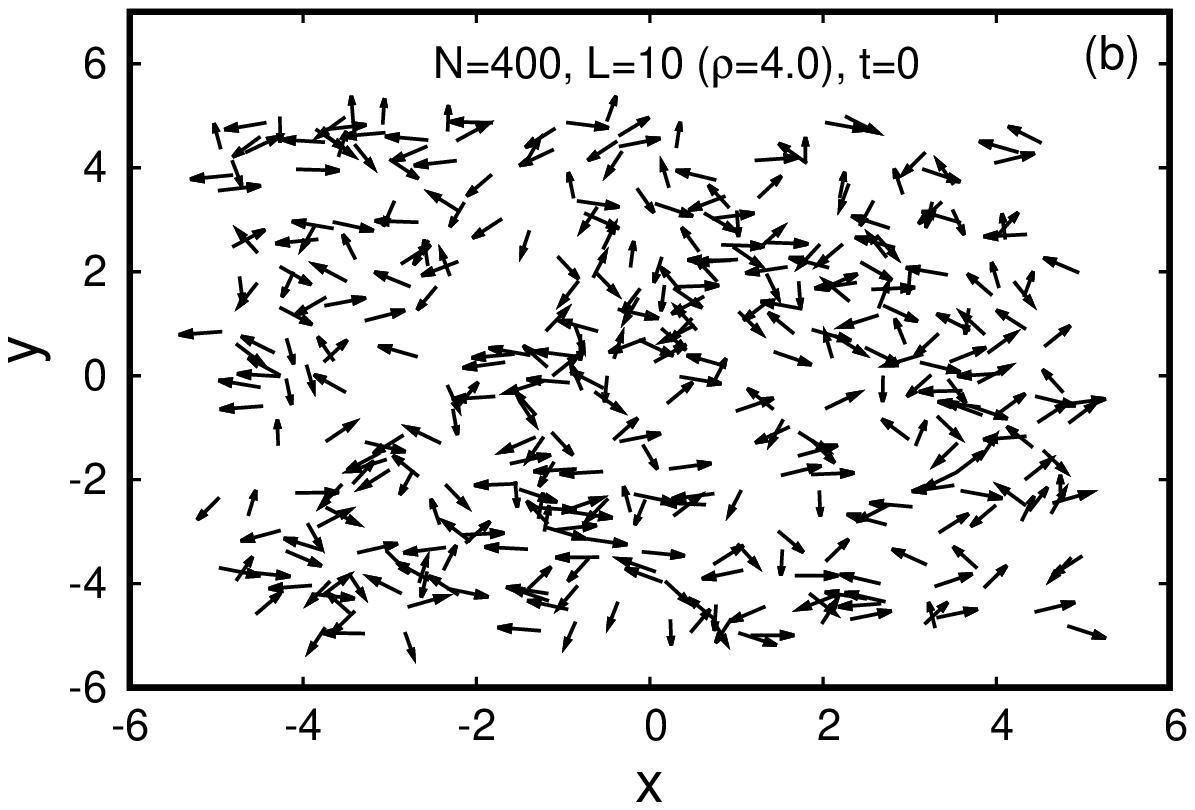}}
\\
\\
\resizebox{7.0cm}{!}{\includegraphics[angle=0]{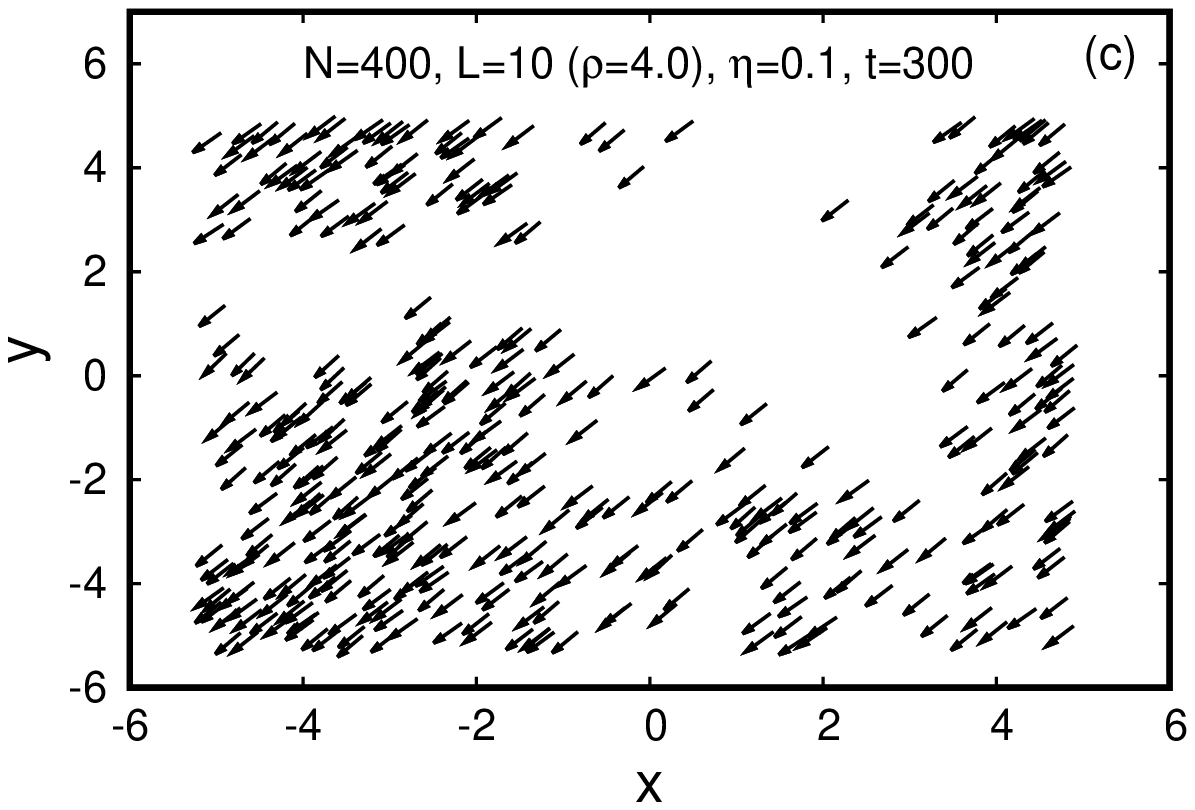}}
\resizebox{7.0cm}{!}{\includegraphics[angle=0]{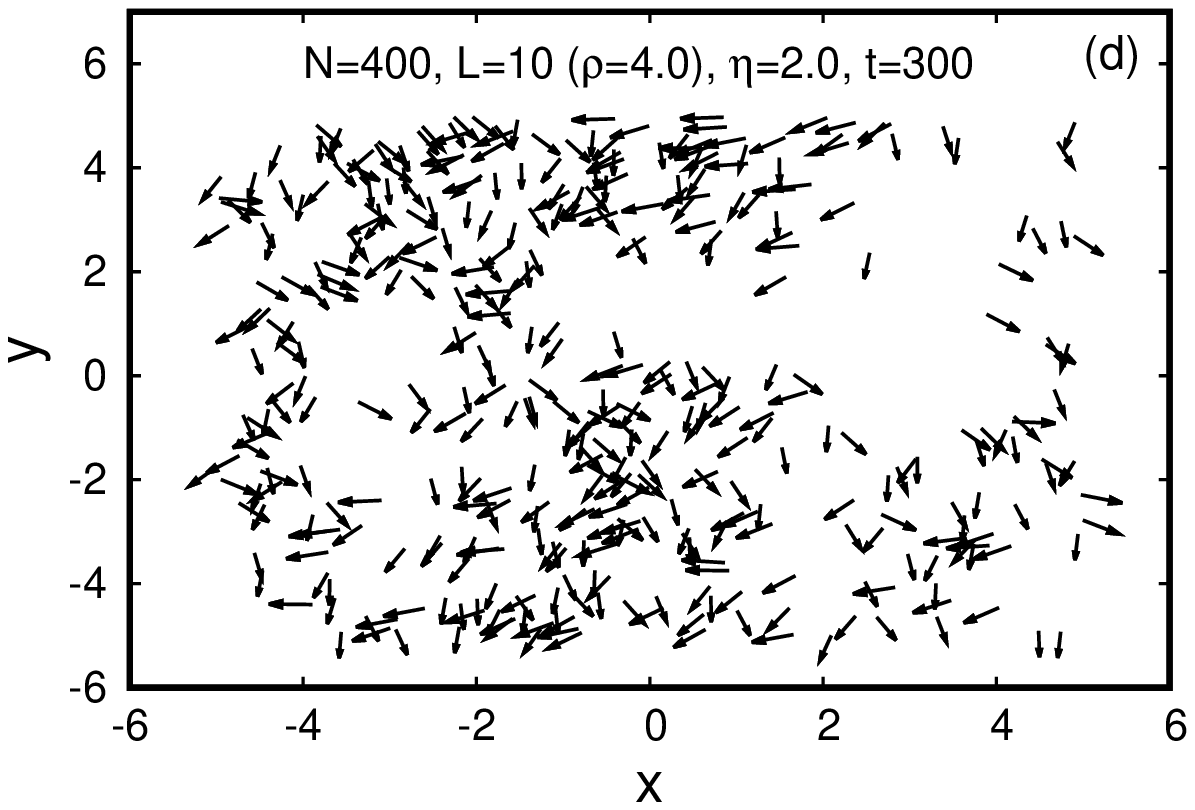}}
          \end{tabular}
\caption{(a) The bold green line denotes the normalised uniform distribution of angles of 
velocity-directions of particles for N=400 at t=0. The blue sharp unimodal peak represents the ordered steady state at t=300 for a particular realisation of the sample at low noise, the corresponding angle distribution at t=300 at higher noise is
represented by the red line which exhibits a smeared peak denoting quasi-randomness;
(b) Velocity morphology of the particles at $t=0$ (completely random);
(c) Velocity morphology of the particles at $t=300$ when the system is subjected to low noise ($\eta=0.1$) (completely ordered);
(d) Velocity morphology of the particles at $t=300$ when the system is subjected to high noise ($\eta=2.0$) (quasi-random).}
\label{fig:noise2}
\end{center}
\end{figure}

We have tried to provide a possible explanation of this contrast behaviour (the time
scale decreases as one approaches the critical noise) of {\it reverse} critical slowing down,
with the help of the time evolution of the profile of the normalised distribution of angles of directions of the particles (Fig. \ref{fig:noise2}a). When the system starts from an initial random configuration (Fig. \ref{fig:noise2}b), the probability distribution of angles is uniform (denoted by bold green line in Fig. \ref{fig:noise2}a), irrespective of low or high noise. But in the low noise case ($\eta=0.1$), the distribution is sharply peaked since all the particles 
become uni-directional (Fig. \ref{fig:noise2}c) upon reaching the steady state (the blue peak in the Fig. \ref{fig:noise2}a denotes low noise steady state case for a single ensemble), whereas when the system is subjected to higher noise ($\eta = 2.0$) the angular distribution exhibits smeared peak (red line in the Fig. \ref{fig:noise2}a) reflecting that the morphology of velocity portrait is quasi-ordered (Fig. \ref{fig:noise2}d), it is neither entirely random nor entirely ordered, yet has a finite non-zero value of the order parameter (Fig. \ref{fig:noise1}a) in the steady state. So, it is quite natural to expect that the time taken to achieve a distribution having a sharp peak (with very small width) from an initial uniform 
distribution (with maximum width $2\pi$) will naturally be larger than the time taken to achieve a distribution having a comparably smeared peak
(with a comparatively larger width). This can be a justifiable 
argument for the decrease of the integrated lifetime with the increase of noise. A video of the simulation can be found in YouTube with the link 
\underline{https://youtu.be/uGOhvfsyc6M}.\\

\vskip 0.5cm

\noindent {\bf 4. Summary:}

\vskip 0.1cm

In this article, we have studied the dynamical evolution of the transient
phases of Vicsek model of flocking, by computer
simulation. 
Although the Vicsek model is widely studied to explore the nature of 
non-equilibrium ordered steady state and the transitions, the behaviours of
the transient phases are somehow overlooked. We have thoroughly investigated
the dynamical evolutions of the morphologies of various 
transient phases designed by the direction of velocity vector
of each agent. T. Vicsek, in his originally proposed model
of flocking (in 1995), mentioned that this model may
map to XY model in the $v \to 0$ limit. This inspired us to search for 
any possible vortex phase (observed in two dimensional XY 
ferromagnet leading to Kosterlitz-Thouless transition\cite{koster}). 
In the low velocity 
limit, 
we have indeed observed the vortex phase. This vortex 
persists (in the low noise limit) with coherent rolling over a considerable scale of time. Moreover, 
for the different values of low density, 
we have also found the colliding phase, where two
groups of oppositely moving agents collide before getting ordered. We have
also proposed a method of characterising this colliding phase and the method
of estimating the lifetime of the colliding phase. A third kind of transient
phase, namely, the multi-domain phase, was also observed. In this phase, the
multiple domains of agents (marked by a variety of net directions 
of velocity vectors)
were found to form in different regions of the entire cell of study.
We have measured the number of domains, strength of each domain and the
lifetime of the multi-domain phase. The integrated lifetime of all these
transient phases is measured statistically and found to follow a scaling
relation with the density, characterised by power law variation with some
exponents.
We have also studied the integrated lifetime as function of the noise.
The integrated lifetime $\tau$ was observed to decrease
as the noise approaches the critical noise from below. 
This is contrary to the critical slowing down, usually observed
in the case of equilibrium critical phenomena.
We have tried to provide a possible
explanation of this unusual behaviour, from the time evolution of the 
distribution of the directions of
velocities as follows: the temporal evolution of the Vicsek dynamics starts from an
initial random (uniformly distributed) configuration of the directions of
motion. This distribution is uniform and has the maximum width ($2\pi$). In the case of
low noise, the final non-equilibrium steady state is an almost uniquely directed motion
of all agents having a sharply peaked distribution of the directions of motion.
On the other hand, for high noise (where the value of the order parameter
is quite low; near the critical noise), the steady state distribution 
of the angles has a
wide spread. So, it is quite natural to expect that, to achieve a very sharp
distribution, the system will take more time. 
As a result, 
the transient phases will be longer lasting for low noise than that for high
noise. 
A live demonstration can be found in
YouTube with the link
\underline{https://youtu.be/uGOhvfsyc6M}, for better understanding.
Extensive numerical studies reported\cite{chate1} the discontinuous nature of this flocking
transition which also supports this finding, since one might not expect the {\it critical
slowing down} in the case of discontinuous transitions.\\

\vskip 0.5cm

\noindent {\bf Acknowledgements:} We would like to
thank Bikas K. Chakrabarti, Surajit Sengupta, Arnab Saha,
Satya N. Majumdar for helpful discussions and useful suggestions. We also thank Sagnik Kayal for a
careful reading of the manuscript and fixing typos and grammatical errors.



\newpage

\newpage

\begin{center} {\bf References} \end{center}
\begin{enumerate}

\bibitem{sriram1} S. Ramaswamy, Annu. Rev. Condens. Matter Physics {\bf 1}
(2010) 323

\bibitem{sriram2} V. Narayan, S. Ramaswamy and N. Menon, Science, {\bf 317}
(2007) 105

\bibitem{sriram3} M. C. Marchetti, J. F. Joanny, S. Ramaswamy, T. B.
Liverpool, J. Prost, M. Rao and R. A. Simha, Rev. Mod. Phys. {\bf 85} (2013)
1143

\bibitem{jtb} I. D. Couzin, J. Krause, R. James, G. D. Ruxton and N. R. Franks,
J. theor. Biol. {\bf 218} (2002) 1

\bibitem{fish} C. K. Hemelrijk, D. A. P. Reid, H. Hildenbrandt and J. T. Padding,
FISH and FISHERIES, {\bf 16} (2015) 511

\bibitem{cell} E. M\'ehes and T. Vicsek, Integr. Biol. {\bf 6} (2014) 831

\bibitem{vicsek} T. Vicsek, A. Czir\'ok, E. Ben-Jacob, I. Cohen and 
O. Schochet, Phys. Rev. Lett. {\bf 75} (1995) 1226

\bibitem{mermin} N. D. Mermin and H. Wagner, Phys. Rev. Lett. {\bf 17} (1966) 1133

\bibitem{revvicsek} T. Vicsek and A. Zafeiris, Physics Reports, {\bf 517}
(2012) 71

\bibitem{ginelli} F. Ginelli, Eur. Phys. J. Special Topics, {\bf 225} (2016)
2099

\bibitem{filella} A. Filella, F. Nadal, C. Sire, E. Kanso and C. Eloy,
Phys. Rev. Lett. {\bf 120} (2018) 198101

\bibitem{manna1} B. Bhattacherjee, S. Mishra and S. S. Manna, Phys. Rev. E
{\bf 92} (2015) 062134

\bibitem{manna2} B. Bhattacherjee, K. Bhattacharya and S. S. Manna,
Frontiers in Physics, {\bf 1} (2014) Article 35

\bibitem{arnab} M. Durve, A. Saha and A. Sayeed, Eur. Phys. J. E (2018)
{\bf 41:} 49

\bibitem{samriddhi} A. Choudhary, D. Venkatraman and S. S. Ray, Europhysics
Letters, {\bf 112} (2015) 24005

\bibitem{solon1} A. P. Solon, H. Chat\'e and J. Tailleur, Phys. Rev. Lett.
{\bf 114} (2015) 068101

\bibitem{solon2} A. P. Solon and J. Tailleur, Phys. Rev. Lett. {\bf 111}
(2013) 078101

\bibitem{solon3} A. P. Solon, Jean-Baptiste Caussin, D. Bartolo, H. Chat\'e
and J. Tailleur, Phys. Rev. E {\bf 92} (2015) 062111

\bibitem{chate1} G. Gr\'egoire and H. Chat\'e, Phys. Rev. Lett. {\bf 92} (2004)
025702

\bibitem{koster} J. M. Kosterlitz and D. J. Thouless, J. Phys. C: Solid 
State Phys. {\bf 6} (1973) 1181

\bibitem{calovi} D. S. Calovi, U. Lopez, S. Ngo, C. Sire, H. Chat\'e and G. Theraulaz, New J. Phys. {\bf 16} (2014) 015026
\end{enumerate}
\end{document}